\documentclass[aps,twocolumn,showpacs,pra,superscriptaddress,floatfix,longbibliography]{revtex4-1}
\usepackage{graphicx,subcaption}%[draft]
\usepackage[intlimits,tbtags]{amsmath}
\usepackage[dvipsnames]{xcolor}
\usepackage{amssymb}
\usepackage{array}
\usepackage{comment}
\usepackage{dcolumn}
\usepackage{bm}
\usepackage{multirow}
\usepackage{siunitx}
\usepackage[version=4]{mhchem}
\usepackage{tabularx}
\usepackage{hyperref}
\usepackage{cleveref}
\usepackage{xr}
\usepackage{soul}

\newcommand{\halle}{Institut f\"ur Physik, Martin-Luther-Universit\"at
  Halle-Wittenberg, D-06099 Halle, Germany}

\newcommand{\mrf}{\multirow{4}{*}}

\begin{document}

\author{Jonathan Schmidt}
\affiliation{\halle} 
\author{Hai-Chen Wang}
\affiliation{\halle} 
\author{Georg Schmidt}
\affiliation{\halle} 
\author{Miguel A. L. Marques} 
\email{miguel.marques@physik.uni-halle.de}
\affiliation{\halle}

\title{Machine Learning guided high-throughput search of non-oxide garnets} %sulfide, selenide, and nitride garnets}

\begin{abstract}
  Garnets, known since the early stages of human civilization, have
  found important applications in modern technologies including
  magnetorestriction, spintronics, lithium batteries, etc. The overwhelming
  majority of experimentally known garnets are oxides, while
  explorations (experimental or theoretical) for the rest of the
  chemical space have been limited in scope.  A key issue is that the
  garnet structure has a large primitive unit cell, requiring an
  enormous amount of computational resources. To perform a
  comprehensive search of the complete chemical space for new garnets,
  we combine recent progress in graph neural networks with high-throughput calculations. We apply the machine
  learning model to identify the potential (meta-)stable garnet
  systems before systematic density-functional calculations to validate the predictions. In this way, we
  discover more than 600 ternary garnets with distances to the convex
  hull below 100~meV/atom with a variety of physical and chemical
  properties. This includes sulfide, nitride and halide garnets. For
  these, we analyze the electronic structure and discuss the
  connection between the value of the electronic band gap and charge
  balance.
\end{abstract}

\maketitle

\section{Introduction}

Garnets can be found throughout the world in diverse geological environments, and have been known since prehistory mainly due to their use in jewelry as gemstones. They are also relatively hard minerals, a property
that makes them useful for a series of industrial applications, such
as in waterjet cutting or as abrasives.

Generally, the garnets crystallize in a cubic structure (space group
$Ia\overline{3}d$) with chemical composition A$_3$B$_2$(B'C$_4$)$_3$,
where the A atoms are located in the $24c$ dodecahedral sites, the B
atoms are in the $16a$ octahedral, and B' atoms occupy the $24d$
tetrahedral sites. In ternary garnets, B and B' sites are occupied by
the same chemical element.  Around
1950~\cite{yoder1951complete,bertaut1956structure,10.1103/PhysRev.110.1311}
some rare-earth garnets, especially yttrium-based materials,
started to attract attention. Those garnets have a general formula of
\ce{RE3B2(BO4)3} where RE stands for rare-earth and B is a 3d magnetic
transition metal (usually iron) or a group IIIA element. Among these,
one of the most used ones is yttrium aluminum garnet (YAG),
\ce{Y3Al2(AlO4)3}, used as a synthetic simulant to diamond due to its
high refractive index ($>1.8$)~\cite{palik1997}. Doped YAGs with other
rare-earth elements have found numerous applications as lasing media
in modern medical laser devices~\cite{luke2019lasers} or in tunable
optical
devices~\cite{10.1016/j.ceramint.2020.01.242,10.1016/j.jallcom.2019.151903,
  10.1002/bio.3108,10.1063/1.111283}.

Other important compounds, with interesting ferrimagnetic properties,
are the rare-earth iron garnets (\ce{RE3Fe2(FeO4)3}, RIG). In the RIG
structure, five Fe atoms occupy two different sublattices, and the
antiferromagnetic coupling between sub-lattices and ferromagnetic
coupling within the sublattice leads to a ferrimagnetic
configuration. RIGs can display a rather high Curie temperature
(around 560~K~\cite{10.1109/TMAG.2011.2180732}), and some systems
exhibit giant magnetorestriction~\cite{10.1016/0304-8853(86)90456-7}.
Moreover, RIGs materials have a band gap with values around 2.6 to
2.9~eV~\cite{10.1134/1.2161290,10.1016/0038-1098(74)90760-1}. Among these materials yttrium iron garnet (YIG) stands out because it has an exceptionally low Gilbert damping. YIG has first been used as bulk material in optical insulators, circulators and Faraday rotators. Since the last two decades, YIG is also more and more frequently used as thin film material for spintronic applications~\cite{wu2013recent} because it allows the transmission of spin currents although being an insulator by itself.  In recent years we have witnessed the attempt to replace yttrium by lanthanides to increase the spin orbit coupling and introduce even Dzyaloshinskii–Moriya interactions in hybrid systems.

Another interesting group of quaternary garnets is the lithium garnets
\ce{LN3M2(LiO4)3}, where LN is a lanthanide and M is either Te, Ta, or
Nb. With partial Li-filling of the invasive positions of tetrahedral
and distorted octahedral sites, the stuffed lithium garnets
(\ce{LN3M2Li2(LiO4)3}) have a promising lithium-ion conductivity and
chemical stability, showing potential as solid electrolytes in
Li-batteries~\cite{10.1016/j.est.2020.102157,10.3389/fchem.2020.00468}.

The many applications of garnets, and of YIG in particular, has increased the need for garnets with different properties. Unfortunately the deposition of garnet thin films with high quality is only possible on garnet substrates due to the special crystal structure~\cite{Schmidt2020,Althammer2018,Serga2010}. Moreover for spintronics applications it would be a huge step forward to be able to pair, for example, thin YIG films in hybrid structures with other conducting films being either metallic or having such a low band gap that reasonable electron conductivity is achievable at room temperature~\cite{Schmidt2020,Althammer2018,Serga2010}. Again to achieve a perfect interface in these structures one would need to create an all-garnet hybrid which is currently prevented by the obvious lack of room temperature conducting or metallic garnets. Such a material class would dramatically extend the applicability of garnet thin films.

However, despite these diverse applications and the technological
relevance of garnets, most of the research in garnets is confined to
oxides~\cite{ye2018deep}, and only a few halides (also called
cryolithionites) are known
experimentally~\cite{10.2138/am.2013.4201}. This is can be easily
understood, as oxides are usually simpler to work with under ambient
experimental conditions. Furthermore, a computational high-throughput
search of new compositions for the garnet prototype is challenging, as
the garnet cubic primitive unit cell contains 80 atoms, which is an
order of magnitude larger than most structure prototypes used in
recent high-throughput searches such as
(double-)perovskites~\cite{CGAT, schmidt2017},
(half)-Heuslers~\cite{doi:10.1021/acs.chemmater.6b02724},
dichalcogenides~\cite{10.1039/D0MA00999G}, etc.

Fortunately, with the aid from state-of-the-art machine learning
techniques, the problem of searching through the entire combinatorial
chemical space can be significantly accelerated. With pre-trained
machine models, we can filter millions of compositions according to
the predicted stability without performing costly density functional
theory (DFT) calculations for all the compositions. Nevertheless, DFT
validation of the stable compounds is still necessary as a
post-processing step.

In the present paper, we followed such a procedure to explore possible
(meta-)stable compounds beyond oxy-garnets. The rest of the paper is
structured as follows. In section \ref{sec:method} we explain the
machine learning model and the computational methods we applied. In
section \ref{sec:results} we present the most interesting
crystal phases we uncovered in our work and discuss the potential
applications of the new proposed compounds. Finally we present our
conclusions and an outlook.

\section{Methods}
\label{sec:method}

\subsection{Machine Learning Model}

In this work we applied crystal graph attention networks, developed
and pretrained in Ref.~\onlinecite{CGAT}, to predict thermodynamically
stable materials. The networks use an attention-based message passing
approach based on the crystal graph representation of the crystal
structure. Replacing the normal distance information that is typically
used as edge-representation in crystal graph networks with solely the
graph distance of the atoms to their neighbors allows for precise
predictions of unrelaxed structures.  As garnets crystallize in a
cubic structure the list of neighbors and consequently the graph
distances are mostly constant throughout the geometry relaxation. This
removes the need to perform predictions with multiple cell constant
ratios.

We study all garnet compositions \ce{A3B5C12} considering all elements
up to Bi, excluding only the rare-gases, and spanning a space of
around 550k possible compounds. Obviously, most of these hypothetical
compounds are highly unstable from the thermodynamic point of view.
As discussed in Ref.~\cite{CGAT}, current machine learning approaches
for the prediction of thermodynamic stability suffer from a
considerable error due to the strong bias present in the existing
datasets. To circumvent this problem, we use the following workflow:
(i)~The machine-learning model is used to predict the distance to the
convex hull of stability for all 550k compounds. At the start we use
the pre-trained machine from Ref.~\cite{CGAT}; (ii)~We perform DFT geometry optimizations to validate all compounds predicted below 200~meV/atom from the hull; (iii)~We add these calculations to a
dataset containing all DFT calculations for garnet systems. (iv)~We use transfer learning to train a new model using this dataset with a training/validation/testing split of 80\%/10\%/10\%. (v)~The cycle is restarted.

In total we repeated the cycle three times. In the first, we performed DFT
calculations for 3320 compounds. The mean absolute error (MAE) of the
initial pre-trained machine was 0.497~eV/atom. This is a very high
value, that was expected as there were very few garnets in the dataset
used in Ref.~\cite{CGAT}, and they spanned a very small chemical
space. In the second cycle we validated 7336 compounds, and the
transfer-learning model performed much better, with a MAE of
0.064~eV/atom. Finally, in the third cycle we computed 3844 materials with
DFT. The final model had a MAE of 0.058~eV/atom, showing that the
tranfer-learning workflow is meaningful and converges quickly.

\subsection{DFT Calculations}

We perform DFT calculations using the package \textsc{vasp} with
PAW~\cite{paw} datasets of version 5.2. and with the
Perdew-Burke-Ernzerhof~\cite{pbe} (PBE) exchange-correlation
functional.  Following the Materials Project~\cite{MP_Jain2013}
recommendations, we use extra on-site corrections for oxides,
fluorides containing Co, Cr, Fe, Mn, Mo, Ni, V, and W. The on-site
corrections are repulsive and correct the d-states by respectively
3.32, 3.7, 5.3, 3.9, 4.38, 6.2, 3.25, and 6.2 eV.  A cutoff of 520\,eV
is applied to the planewaves, and $\Gamma$-centered $k$-point grids
with a uniform density of 1000~k-points per reciprocal atom are used
to sample the Brillouin zone. For geometry optimizations all forces
are converged to less than 0.005\,eV/\AA.

All calculations are performed with spin-polarization, starting from a
ferromagnetic ground-state as is customary in high-throughput
searches. Unfortunately, this means that, in most cases,
antiferromagnetic or ferrimagnetic systems will converge to an
incorrect ferromagnetic ground-state. This is important, in our
context, particularly for ferrimagnetic garnets having a 3d transition
metal such as Fe, Ni, Co, Cr, Mn, or V in the B position. We note that
not only the spin-state of these garnets but also of anti- and
ferrimagnetic systems on the convex hull are treated
incorrectly. Consequently, the estimation of $E_\text{hull}$ for
ferrimagnetic garnets is far less accurate than for non-magnetic
ones. For example, the experimentally know \ce{Gd3Fe5O12} is predicted
to be more than 1~eV above the hull according to Materials Project
database~\cite{MP_Jain2013}, a value that is certainly grossly
overestimated.

To properly estimate $E_\text{hull}$ for the ferrimagnetic garnets
would require obtaining the correct magnetic ordering for a portion
of the convex hull as well as for the garnets. This is a complex and
computationally expensive task, that is well beyond the scope of this
work. Therefore, we made the choice to restrict our discussion to
systems not containing the 3d metals mentioned above in the B-site.

It is well-known that the electronic band gaps calculated with the PBE
functional are severely underestimated~\cite{benchmark1}. Therefore,
to obtain a more reliable estimation of this important physical
property we performed calculations with the modified Becke-Johnson
(mBJ) approximation~\cite{PhysRevLett.102.226401}, as this is by now
recognized as one of the most accurate functional for this
task~\cite{benchmark2}.

 To calculate the averaged carrier effective masses from the interpolated eigenvalues we follow the approach of Ref.~\cite{HautierEff}. Considering a temperature of 300~K, we deduce the chemical potential required to reach a reference carrier concentration (10$^{18}$~cm$^{-3}$) by using \textsc{BoltzTraP2}. 
We use a $k$-point mesh with a regular density of 2000~$k$-points per reciprocal atom and interpolated the calculated eigenvalues using \textsc{BoltzTraP2}~\cite{Boltztrap,madsen2018boltztrap2}.
 The calculated averaged carrier effective masses can be seen as the intrinsic tendency for creating mobile charge carriers in materials~\cite{HautierEff}.

%explanation for effective mass calculation in here https://static-content.springer.com/esm/art%3A10.1038%2Fncomms3292/MediaObjects/41467_2013_BFncomms3292_MOESM776_ESM.pdf 

\section{Results and Discussion}
\label{sec:results}

\subsection{Stability and Chemistry}

\begin{figure}[ht]
    \centering
    \begin{subfigure}[t]{0.90\columnwidth}
        \includegraphics[height=5cm]{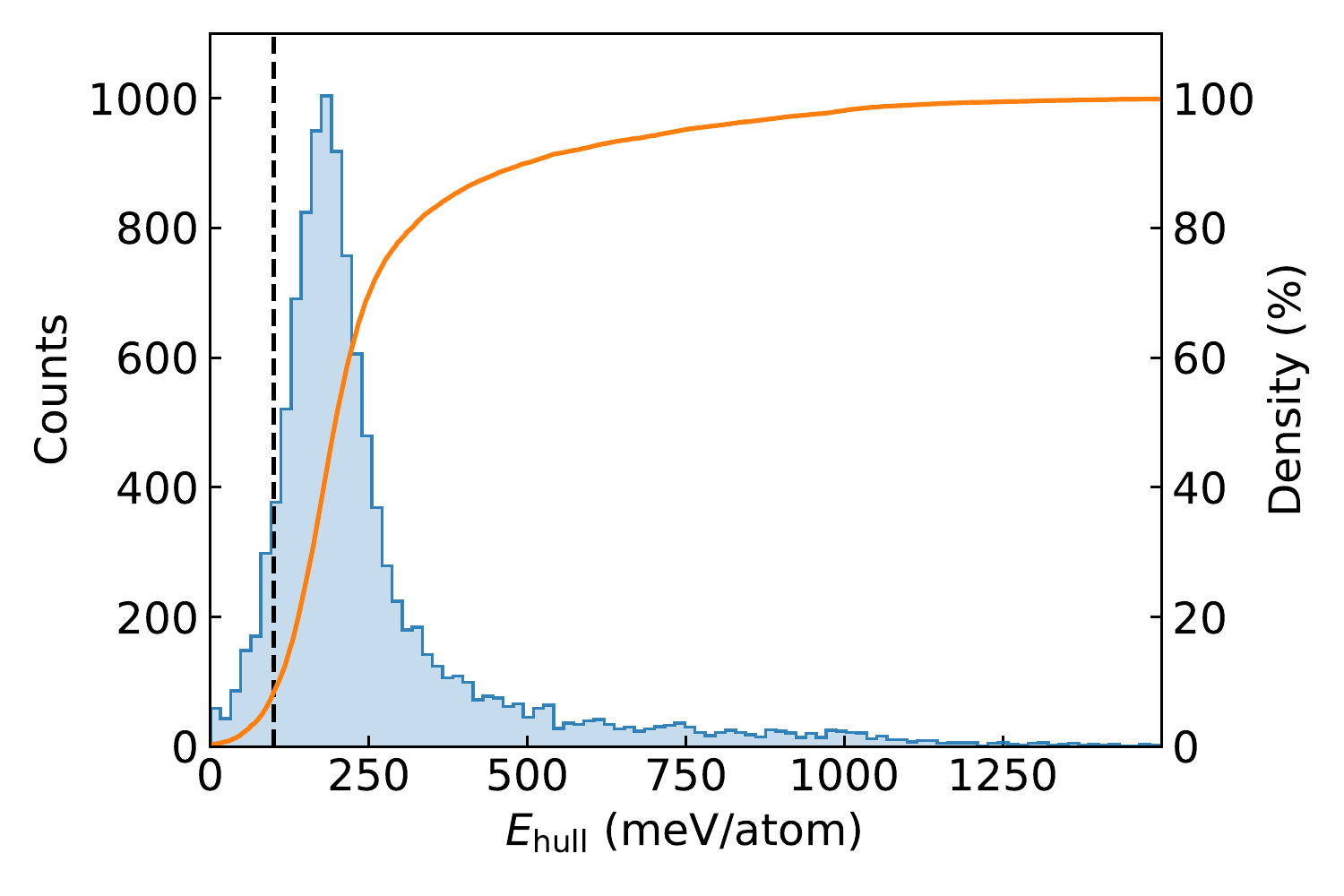}
        \caption{\label{fig:Ehull_hist_all}}
    \end{subfigure}
    \begin{subfigure}[t]{0.90\columnwidth}
        \includegraphics[height=5cm]{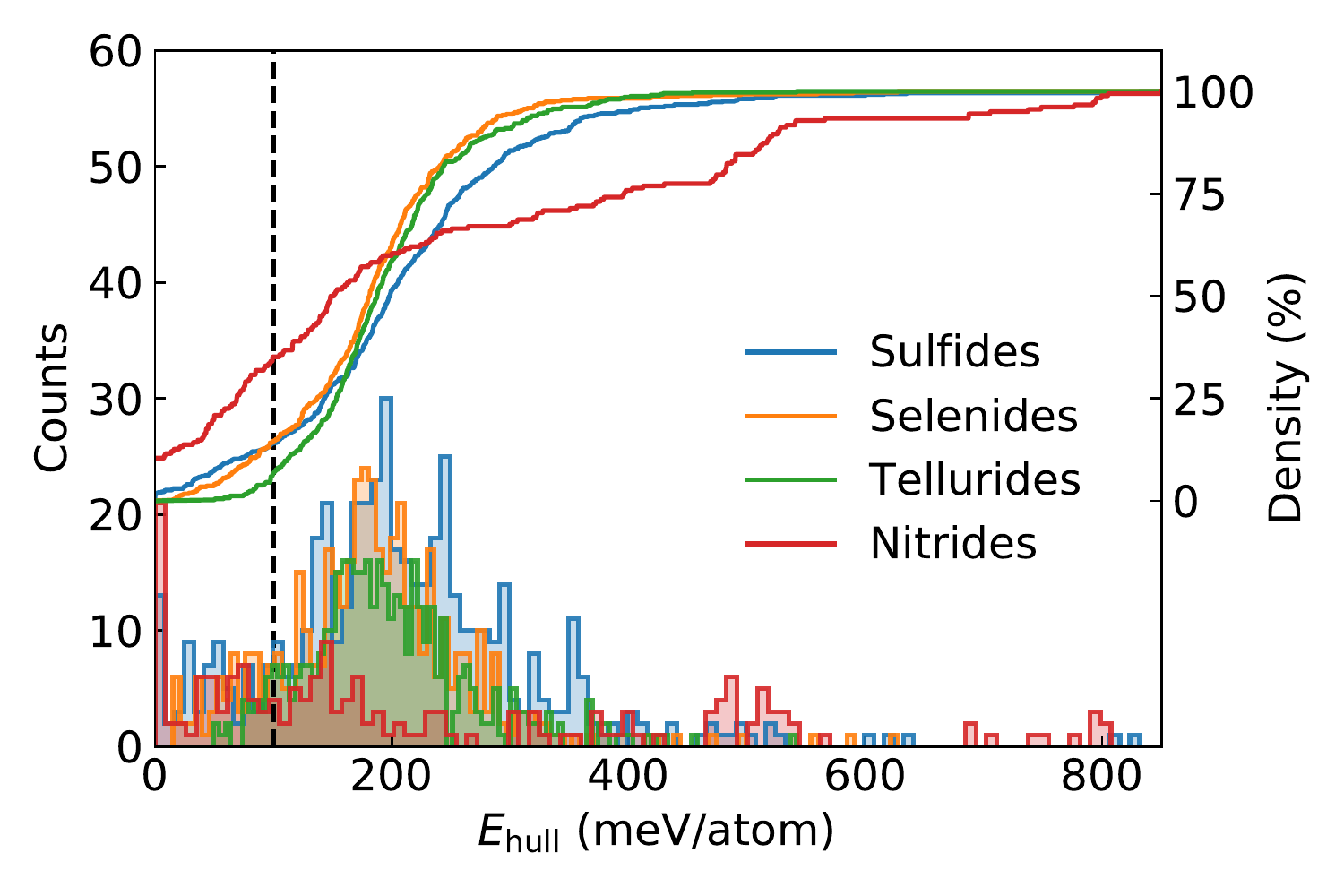}
        \caption{\label{fig:Ehull_hist_sseten}}
    \end{subfigure}
    \begin{subfigure}[t]{0.90\columnwidth}
        \includegraphics[height=5cm]{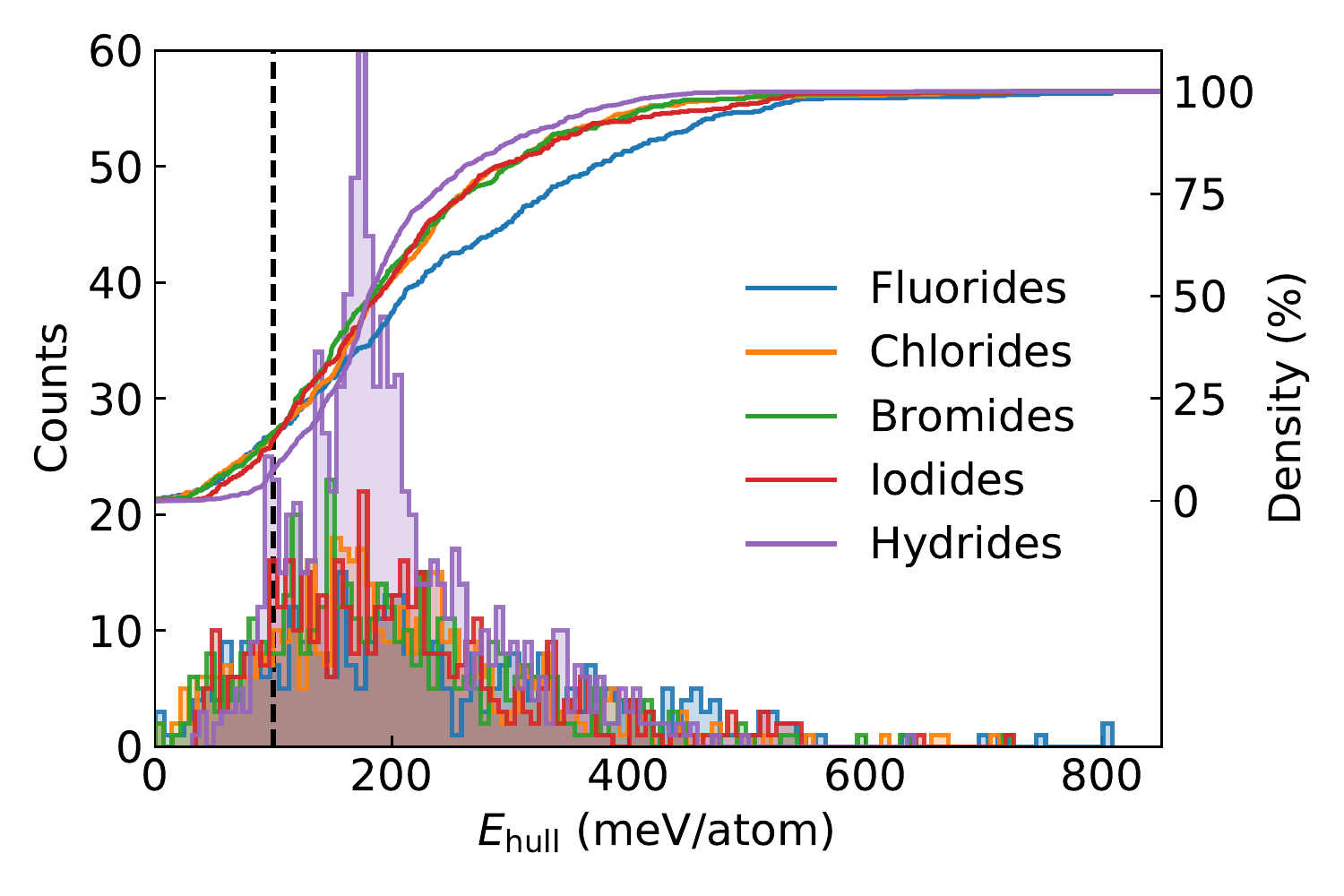}
        \caption{\label{fig:Ehull_hist_halides}}
    
    \end{subfigure}
    \caption{The distribution of the distance to the convex hull ($E_\text{hull}$) for (a)~all calculated compositions, (b)~halides, and (c)~chalcogenides/nitrides. The cutoff to filter (meta-)stable systems (100~meV/atom) is shown as dashed vertical line. }
    \label{fig:Ehull_hist}
\end{figure}

\begin{table}[]
    \centering    
    \caption{Number(meta-)stable systems ($N_\text{stable}$) below 100~meV/atom from the convex hull of thermodynamic stability and high-throughput success rate ($R$) for each category.}
    \label{tab:numbers}
    \begin{tabular*}{\columnwidth}{@{\extracolsep{\fill} } l r r}
         Category & $N_\text{stable}$ & $R$ (\%)   \\
         \hline\noalign{\vskip 2mm}
        Sulfides    & 70  & 14  \\\noalign{\vskip 1mm} 
        Selenides   & 68  & 15  \\\noalign{\vskip 1mm}
        Tellurides  & 28  & 7   \\\noalign{\vskip 1mm}
        Nitrides    & 64  & 35  \\\noalign{\vskip 1mm}
        Fluorides   & 62  & 17  \\\noalign{\vskip 1mm}
        Chlorides   & 68  & 16  \\\noalign{\vskip 1mm}
        Bromides    & 72  & 17  \\\noalign{\vskip 1mm}
        Iodides     & 68  & 15    \\\noalign{\vskip 1mm}
        Hydrides    & 69  & 8    \\\noalign{\vskip 1mm}
        Total       & 569 & 14  \\\noalign{\vskip 1mm}
        \hline
    \end{tabular*}
\end{table}

In total the machine-learning model predicted around 12300
compositions below 200~meV/atom from the convex hull of thermodynamic
stability that were not present in the Materials Project
database~\cite{MP_Jain2013} nor in the Inorganic Crystal Structure
Database~\cite{ICSD} (ICSD). All these calculations can be downloaded 
from the Materials Cloud repository~\cite{linktodata}.

After the DFT validation calculations we
re-evaluated the distance to the convex hull ($E_\text{hull}$) of
these candidates using the much more complete convex hull of
Ref.~\cite{CGAT}.  The histogram of the values $E_\text{hull}$ is
shown in Fig.~\ref{fig:Ehull_hist_all}. We also separate the systems
into sulfides, selenides, tellurides, nitrides, chlorides, bromides,
iodides, and hydrides. These comprise the majority of all systems
found. Most of the candidates have an $E_\text{hull}$ larger than
100~meV/atom, but there are still more than one thousand (about 9\%)
compositions below this threshold. The high-throughput success rate,
that we define by the number of compounds that are within 100~meV/atom
from the convex hull divided by the total number of DFT calculations,
stood at 14\%, with a maximum of 35\% for nitrides and a minimum of
8\% for hydrides. These numbers prove the efficiency of our
machine-learning assisted high-throughout search.  The histogram of
$E_\text{hull}$ for these categories as well as all calculated
compositions are shown in Fig.~\ref{fig:Ehull_hist}.

The distribution of $E_\text{hull}$ for all systems follows the
typical skewed Gaussian with the peak located at around 200~meV/atom and a
fat tail that extends beyond 1~eV/atom, in agreement with the MAE
errors for our machine-learning models.  Individual distributions for
chalcogenides are also skewed Gaussians peaking at around 200~meV/atom. The amount of (meta-)stable compounds decreases from
sulfides to tellurides, which is also expected as this is the common
trend of stability for the chalcogenides. Unlike the situation for
chalcogenides, the distribution curve for nitrides has multiple peaks,
and shows that there are plenty of potentially stable nitride garnets.
For halides and hydrides the histograms are again skewed Gaussian
similar to those of chalcogenides, but there is no clear trend in what
concerns stability across the group. For hydrides, the total number of
systems is much larger, but a lower percentage of them are
(meta-)stable compared to the halides.

The total number of (quasi-)stable systems for each category is listed
as Table~\ref{tab:numbers}. A full list of the systems can be found in
Table~1 of the Supplementary Information (SI). We also selected a dozen
of them to analyze more closely in Table~\ref{tab:sngarnets}. Later
discussions will mainly focus on these systems.

\begin{table*}[tbh]
    \centering
    \caption{The experimentally known oxy-garnets (not including a 3d
      metal in the B site), their ICSD ID, Materials Project ID, and
      predicted meta-stable counterparts with different C anions. The
      distance to the hull calculated with DFT is in parentheses (in
      meV/atom).}
    \begin{tabular}{l l l c c c c c}
    Formula & ICSD ID & MP ID & Counterparts                                                                     \\
    \hline\\[-3mm]
    \multirow{4}{*}{\ce{Y3Al5O12}}
                   &    20090, 41144, 41145, 67102, 67103,       & \mrf{mp-3050}& \mrf{\ce{Y3Al5S12}(-4); \ce{Y3Al5Se12}(36); \ce{Y3Al5Te12}(84) }  \\
                   &    93634, 93635, 170157, 170158, 236589,    &                                    \\
                   &    280104, 17687, 17688, 17689, 17690,      &                                    \\
                   &    74607, 31496                             &                                     \\
    \ce{La3Al5O12} &                                             & mp-780432  & \ce{La3Al5S12}(-4); \ce{La3Al5Se12}(19); \ce{La3Al5Te12}(49) \\
    \ce{Eu3Al5O12} &    245326                                   & mp-21757   &  \ce{Eu3Al5S12}(39); \ce{Eu3Al5Se12}(93)     \\
    \ce{Tb3Al5O12} &    33602                                    & mp-14387   &  \ce{Tb3Al5S12}(-4); \ce{Tb3Al5Se12}(29); \ce{Tb3Al5Te12}(79)  \\
    \ce{Er3Al5O12} &    170147, 280606, 170146, 62615            & mp-3384    &  \ce{Er3Al5S12}(6); \ce{Er3Al5Se12}(50); \ce{Er3Al5Te12}(98) \\
    \ce{Gd3Al5O12} &    192184                                   & mp-14133   & \ce{Gd3Al5S12}(-16); \ce{Gd3Al5Se12}(21); \ce{Gd3Al5Te12}(66) \\
    \ce{Ho3Al5O12} &    409390, 33603                            & mp-14388   & \ce{Ho3Al5S12}(-1); \ce{Ho3Al5Se12}(44); \ce{Ho3Al5Te12}(92) \\
    \ce{Lu3Al5O12} &    259144, 17789, 182354                    & mp-14132   & \ce{Lu3Al5S12}(28); \ce{Lu3Al5Se12}(70)\\
    %\ce{Yb3Al5O12} &    170159, 170160, 280607                   & mp-3800    & --                                     \\
    \ce{Y3Ga5O12}  &    80148, 14343, 185862, 23852              & mp-5444    & \ce{Y3Ga5S12}(47); \ce{Y3Ga5Se12}(81)\\
    \ce{La3Ga5O12}  &                                            & mp-780561  & \ce{La3Ga5S12}(36); \ce{La3Ga5Se12}(51); \ce{La3Ga5Te12}(100)\\
    \ce{Tb3Ga5O12} &    20831, 84875, 184934                     & mp-5965    & \ce{Tb3Ga5S12}(37); \ce{Tb3Ga5Se12}(75)\\
    \ce{Sm3Ga5O12} &    9236, 84873, 291192                      & mp-5800    & \ce{Sm3Ga5S12}(29); \ce{Sm3Ga5Se12}(64); \ce{Sm3Ga5Te12}(93)\\
    \ce{Nd3Ga5O12} &    84872                                    & mp-15239   & \ce{Nd3Ga5S12}(26); \ce{Nd3Ga5Se12}(54); \ce{Nd3Ga5Te12}(96)\\
    \ce{Gd3Ga5O12} &    9237, 37145, 192181, 84874, 184931       & mp-2921    & \ce{Gd3Ga5S12}(32); \ce{Gd3Ga5Se12}(67); \ce{Gd3Ga5Te12}(93)\\
    %\ce{Yb3Ga5O12} &    23851                                    & mp-562499  & --                                    \\
    \ce{Lu3Ga5O12} &    23850                                    & mp-14134   & \ce{Lu3Ga5S12}(74)\\
    \ce{Dy3Ga5O12} &    409391                                   & mp-15576   & \ce{Dy3Ga5S12}(49); \ce{Dy3Ga5Se12}(81)     \\
    \ce{Er3Ga5O12} &    9238                                     & mp-12236   & \ce{Er3Ga5S12}(55); \ce{Er3Ga5Se12}(95)      \\
    \ce{Ho3Ga5O12} &    409390                                   & mp-15575   &  \ce{Ho3Ga5S12}(52); \ce{Gd3Ga5Se12}(89)      \\

    %\multirow{3}{*}{\ce{Y3Fe5O12}}
    %               &    80139, 88502, 88504, 88506, 2012,        & \mrt{mp-4707}    & \mrt{--}                                           \\
    %               &    14342, 29222, 60167, 29235, 173997,      &                                                    \\
    %               &    28561, 23855                             &                                                     \\
    %\ce{Gd3Fe5O12} &    15456                                    & mp-557370  &    --    \\
    %\ce{Tb3Fe5O12} &    9233, 22320, 33602                       & mp-19710   &    --    \\
    %\ce{Pr3Fe5O12} &    248013, 422690                           & mp-1197438 &    --    \\
    %\ce{Yb3Fe5O12} &    23854                                    & mp-562055  &    --  \\
    %\ce{Sm3Fe5O12} &    23857                                    & mp-14139   &    --    \\
    %\ce{Nd3Fe5O12} &    260556, 422691                           & mp-1201849 &    --     \\
    %\ce{Er3Fe5O12} &    71464                                    & mp-560103   &   --     \\
    %\ce{Eu3Fe5O12} &    9232                                     & mp-555758   &   --    \\
    %\ce{Dy3Fe5O12} &    23856                                    & mp-21974   &    --   \\
    %\ce{Lu3Fe5O12} &    23853                                    & mp-559750   &   --   \\
    %\ce{Si3Fe5O12} &    77434, 27377, 161139                     & mp-555755   &   --    \\
    %\ce{Si3Mn5O12} &    27382, 86935, 86936                      & mp-5802    &    --    \\
    \end{tabular}
    \label{tab:expgarnets}
\end{table*}

From our calculations we recover the majority of the oxide (and
halide) garnets that are already known experimentally, but we also
obtain a wealth of different compounds not present in available
databases.  Many of these systems are oxides, that have been the
subject of a recent high-throughput search~\cite{ye2018deep} with
results similar to ours. Interestingly, we also find a wealth of other
chalcogenides, nitrides, halides, and even hydrides as shown in
Figs.~\ref{fig:Ehull_hist_sseten},\ref{fig:Ehull_hist_halides} and in
Table\ref{tab:numbers}.

From the stable compounds, several are closely related to the oxide
garnets by the chemical substitution of oxygen by another
chalcogen. These garnets along with their counterparts are presented
in Table~\ref{tab:expgarnets}. The existence of such compounds is
expected due to the chemical similarity among chalcogens.  In those
chalcogenides, the dodecahedral sites (site A in \ce{A3B5S12}) are
mostly occupied by rare earth elements, and according to the element
occupying the octahedral and tetrahedral sites (site B), the
chalcogenides can be further divided into several categories. The
numbers of (meta-)stable systems for each category decrease in the
following order: occupying B with group IIIA elements (Al, Ga, In,
Tl), group IVA elements (Ge, Sn, Pb), group VA elements (As, Sb, Bi),
and transition metals (Ag, Cu, Sc, Ti). The preference of group IIIA
elements for site B can be understood by simple charge compensation
arguments. The most common oxidation state of the chalcogens is $-2$,
while the rare earth elements in sites A are $+3$: the composition
reaches the ``balanced'' or ``compensated'' state if B is in the
oxidation state $+3$. Moreover, for balanced
A$^\text{III}_3$B$^\text{III}_5$C$^\text{-II}_{12}$ chalcogenide compositions,
one would expect the compounds to be semiconductors. This is indeed
what we find (see an example in
section~\ref{sec:electronic_structures}).

Besides the chalcogenides, we discover 64 stable nitrides as seen in
Table~\ref{tab:numbers}. As discussed above, to reach a balanced
oxidation state, elements with higher oxidation state should be
favored to combine with $-3$ oxydation state of nitrogen. Indeed for
nitride garnets \ce{A3B5N12}, the position A is mostly occupied by
$+2$ or $+3$ chemical elements, and for position B the majority of
meta-stable systems have elements with oxidation state $+6$ (Mo and
W). Although a $+6$ element is required to achieve a balanced state,
we also find that relatively stable compounds are possible for $+5$
(Nb and Ta), and $+7$ (Re). We could argue that the balanced nitrides
should be semiconducting. However, due to the low electronegativity of
nitrogen the gap may close (see one such example in section~\ref{sec:electronic_structures}).

We can also identify several halides and hydrides from
Figure~\ref{fig:Ehull_hist_halides}. However, halogens (and hydrogen
in hydrides) have an oxidation state of $-1$, which makes it more
difficult to reach a balanced oxidation state with the \ce{A3B5C12}
stoichiometry. One viable scheme is with an A element that is $+1$,
while the B elements in $16a$ and $24d$ Wycoff positions are
respectively $+1$ and $+3$, i.e. having the form of
A$^\text{I}_3$B$^\text{I}_2$(B$^\text{III}$C$^\text{-I}_4$)$_{3}$. Chemical
elements exhibiting both $+1$ and $+3$ oxidation states are quite
rare. Nevertheless, we still discovered some meta-stable
semiconducting halide garnets, for example \ce{K3In5F12}.

There are other compositions that do not belong to any of the
discussed classes, such as for example \ce{Sr12Zn3H5}. Most of them
are ``inverted''-garnets, i.e. with cations instead of anions occupying
the C-sites, and have a comparatively higher $E_\text{hull}$ than
regular garnets. Furthermore, only a few anti-garnets with the A and B
sites both occupied by the chemical elements of the nitrogen group
could potentially become oxidation state balanced.

We have to again emphasize that, in order to form (meta-)stable or
insulating/semiconducting compounds, charge compensation is neither a
necessary nor a sufficient condition, and we find many exceptions in
Table S1 in the Supplementary Information. However, it gives us a
simple, intuitive argument to understand why a system is stabilized or
has an electronic band gap. Furthermore, we have to keep in mind that
uncompensated systems might be further stabilized through defects,
such as vacancies. To simplify our discussion, we leave such
possibilities to future works, and focus in the following on (meta-)stable regular
garnets systems which could have balanced charges, specifically
chalcogenides (except oxides), halides, hydrides, and nitrides.

\begin{table}[tbh]
    \centering
    \caption{The calculated lattice constant ($a$, in \AA), band gap
      calculated with PBE (Gap$^\text{PBE}$ ) and MBJ
      (Gap$^\text{MBJ}$) functional (in unit of eV)), distance to the
      convex hull ($E_\text{hull}$ in meV/atom), effective electron
      ($m_\text{e}^*$) and hole ($m_\text{h}^*$) masses (in unit of $m_\text{e}^0$), for
      some selected (meta-)stable sulfide and nitride garnets, data
      for \ce{Y3Al5O12} is also listed for comparison.}
    \begin{tabular}{l c c c c c c}
      Formula & $a$ & Gap$^\text{PBE}$ & Gap$^\text{MBJ}$  & $E_\text{hull}$  & $m_\text{e}^*$ & $m_\text{h}^*$ \\
      \hline \\[-3mm]
      \ce{Y3Al5O12}  & 12.125 &  4.53   &    6.12    &     0     & 1.3 &    6.8     \\
      \ce{Y3Al5S12}  & 14.932 &  2.08   &    3.00    &     0     & 0.7 &    1.5     \\
      \ce{Y3Ga5S12}  & 15.073 &  1.32   &    2.45    &     46    & 0.5 &    3.2   \\ 
      \ce{Y3In5S12}  & 15.709 &  1.35   &    2.38    &     46    & 0.5 &    2.6     \\
      
      \ce{Y3Al5Se12} & 15.752 &  1.46   &    2.15    &     36    & 0.5 &    1.5     \\
      \ce{Y3Al5Te12} & 17.120 &  0.60   &    1.07    &     84    & 0.3 &    2.7     \\
      
      \ce{Y3Ge5S12}  & 15.324 &  0.00   &    0.00    &     81    & --  &    --      \\
      
      \ce{Ca3W5N12}  & 12.928 &  0.93   &    1.57    &     0     & 0.9 &   3.2       \\
      \ce{Ca3Re5N12} & 12.896 &  0.00   &    0.00    &     9     & --  &    --      \\
      \ce{La3Nb5N12} & 13.297 &  0.00   &    0.00    &     99    & --  &    --          \\
      
      \ce{K3In5F12}  & 14.898 &  2.41   &    3.39    &     59    & 0.8 &   6.1     \\
      \ce{K3In5I12}  & 19.977 &  0.00   &    0.00    &     61    & --  &    --      \\
      
      \ce{Mg3Rh5H12} & 11.008 &  0.00   &    0.00    &     51    & --  &    --      \\
      \ce{Y3Rh5H12}  & 11.581 &  0.00   &    0.00    &     92    & --  &    --    \\
    \end{tabular}
    \label{tab:sngarnets}
\end{table}

\subsection{Electronic Structure}
\label{sec:electronic_structures}

\begin{figure}[!th]
   \centering
   \begin{subfigure}[t]{5.15cm}
     \includegraphics[width=5.15cm]{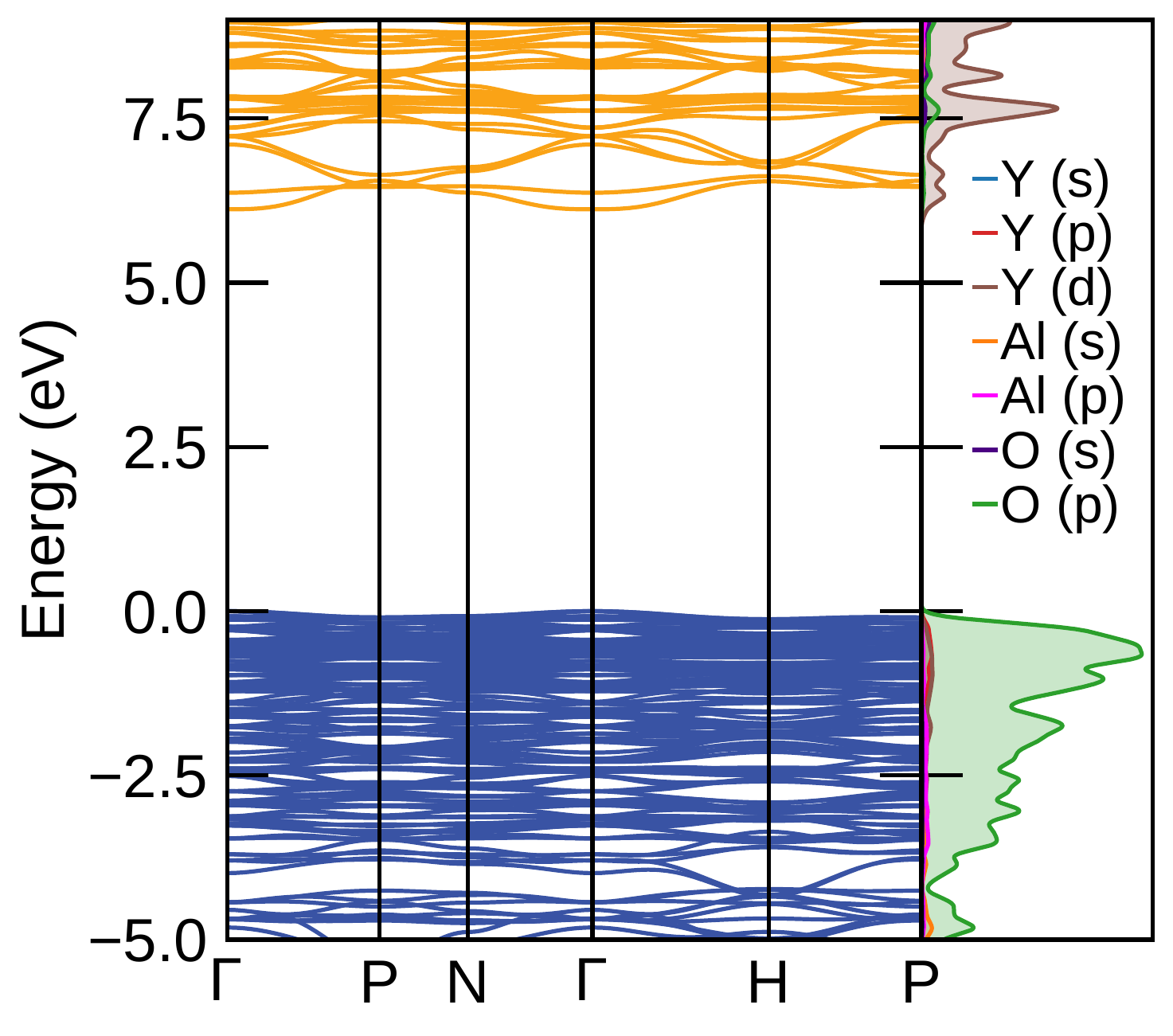} 
     \caption{\label{fig:bands_y3al5o12}~\ce{Y3Al5O12}}
   \end{subfigure}
   \begin{subfigure}[t]{5.15cm}
     \includegraphics[width=5.15cm]{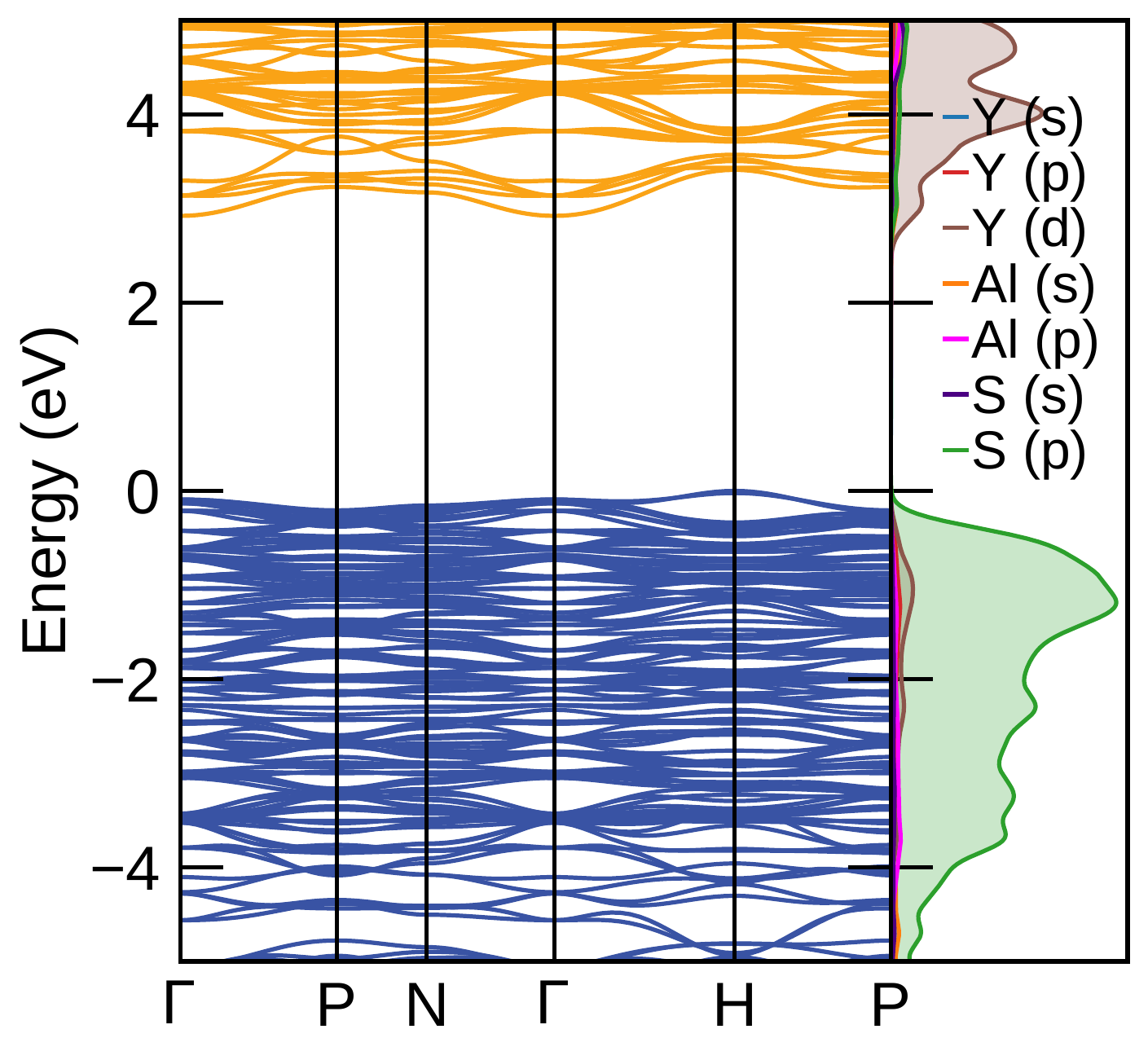}
     \caption{\label{fig:bands_y3al5s12}~\ce{Y3Al5S12}}
   \end{subfigure}
   \begin{subfigure}[t]{5.15cm}
     \includegraphics[width=5.15cm]{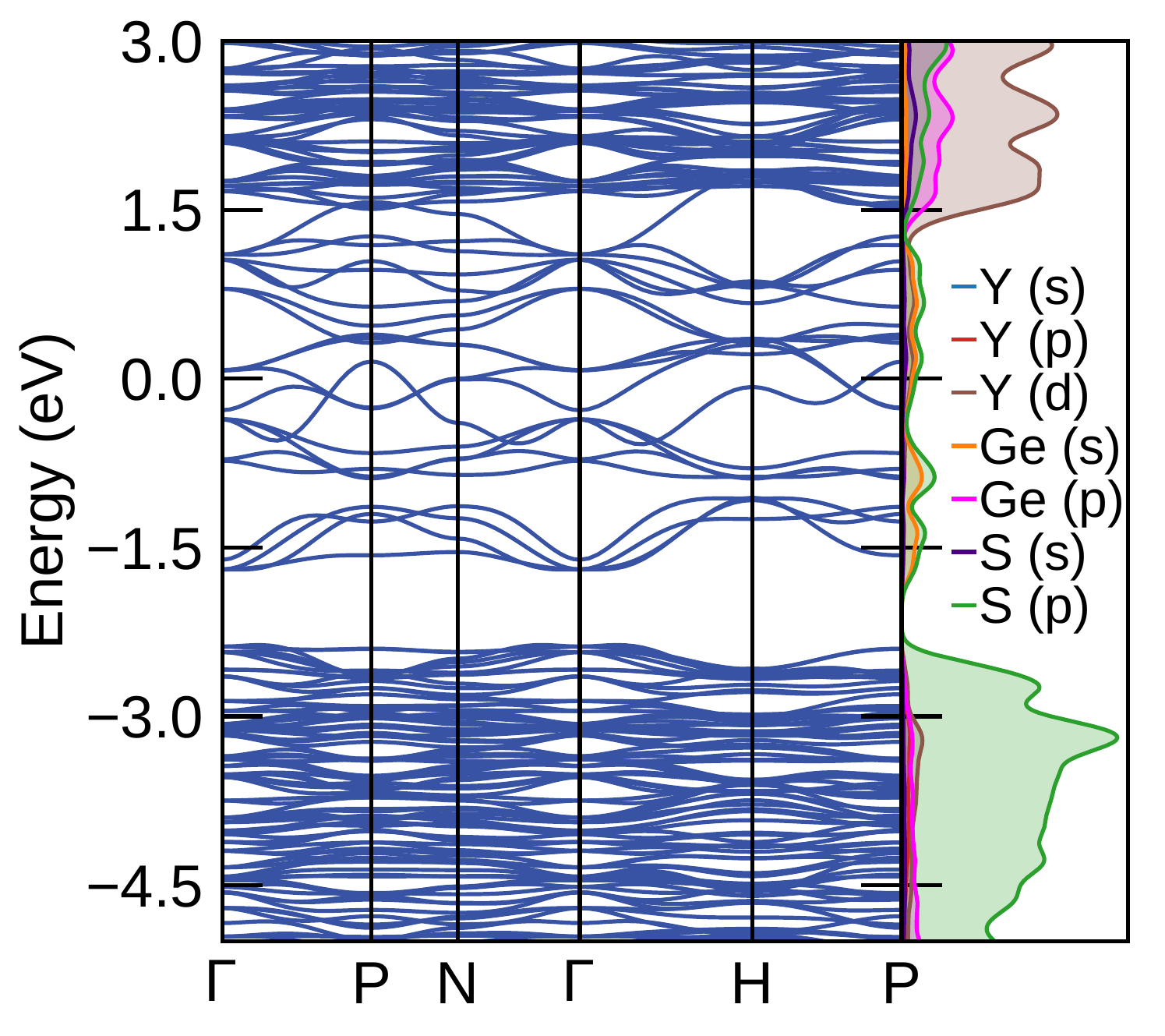}
     \caption{\label{fig:bands_y3ge5s12}~\ce{Y3Ge5S12}}
   \end{subfigure}
   \begin{subfigure}[t]{5.15cm}
     \includegraphics[width=5.15cm]{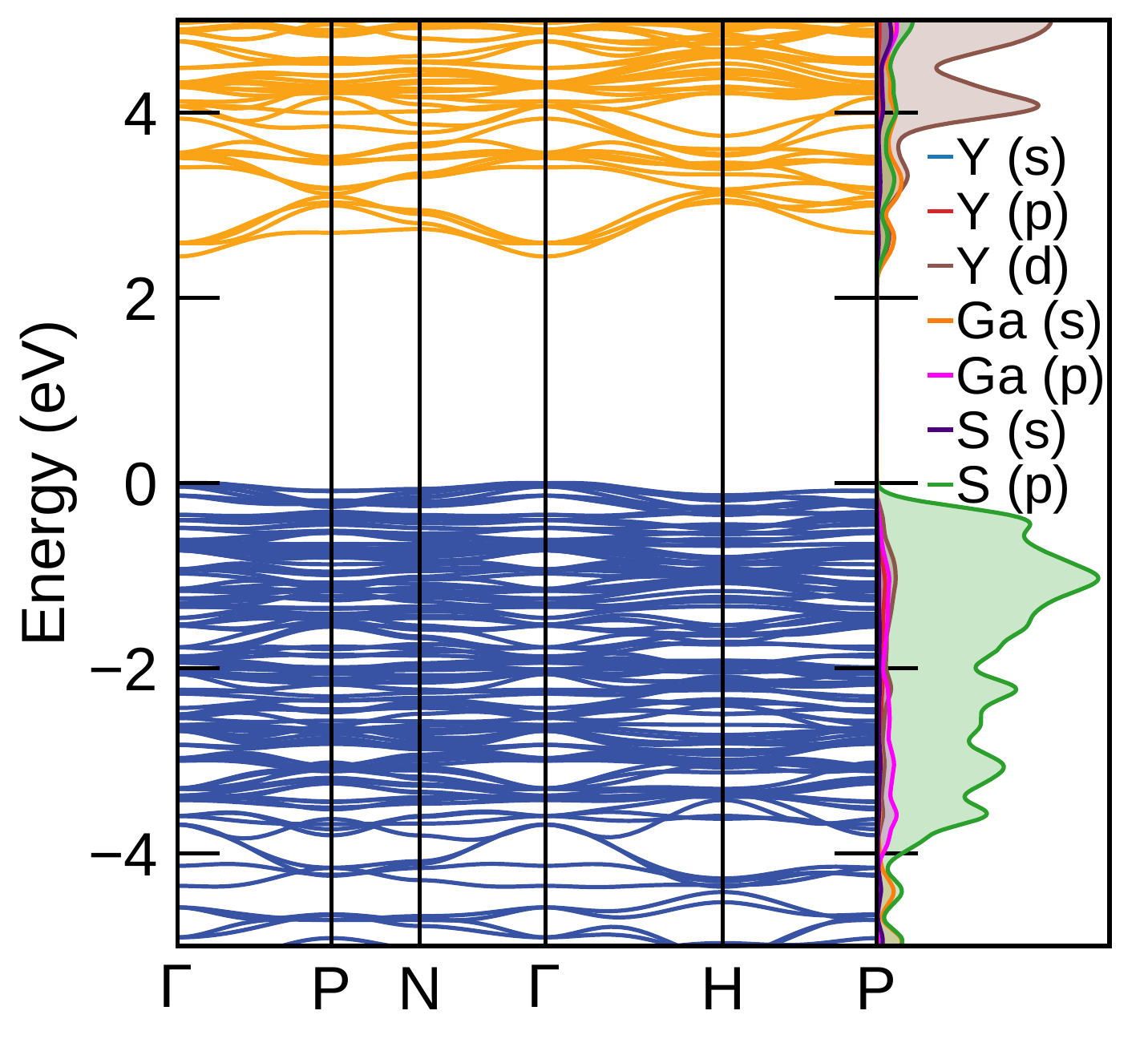} 
     \caption{\label{fig:bands_y3ga5s12}~\ce{Y3Ga5S12}}
   \end{subfigure}

   \caption{Calculated mBJ electronic band structures for selected
     chalgogenide garnets. The Fermi level is set at zero.}
   \label{fig:Band_chalcogenides}
\end{figure}

\begin{figure}[!th]
  \begin{subfigure}[t]{5.15cm}
    \includegraphics[width=5.15cm]{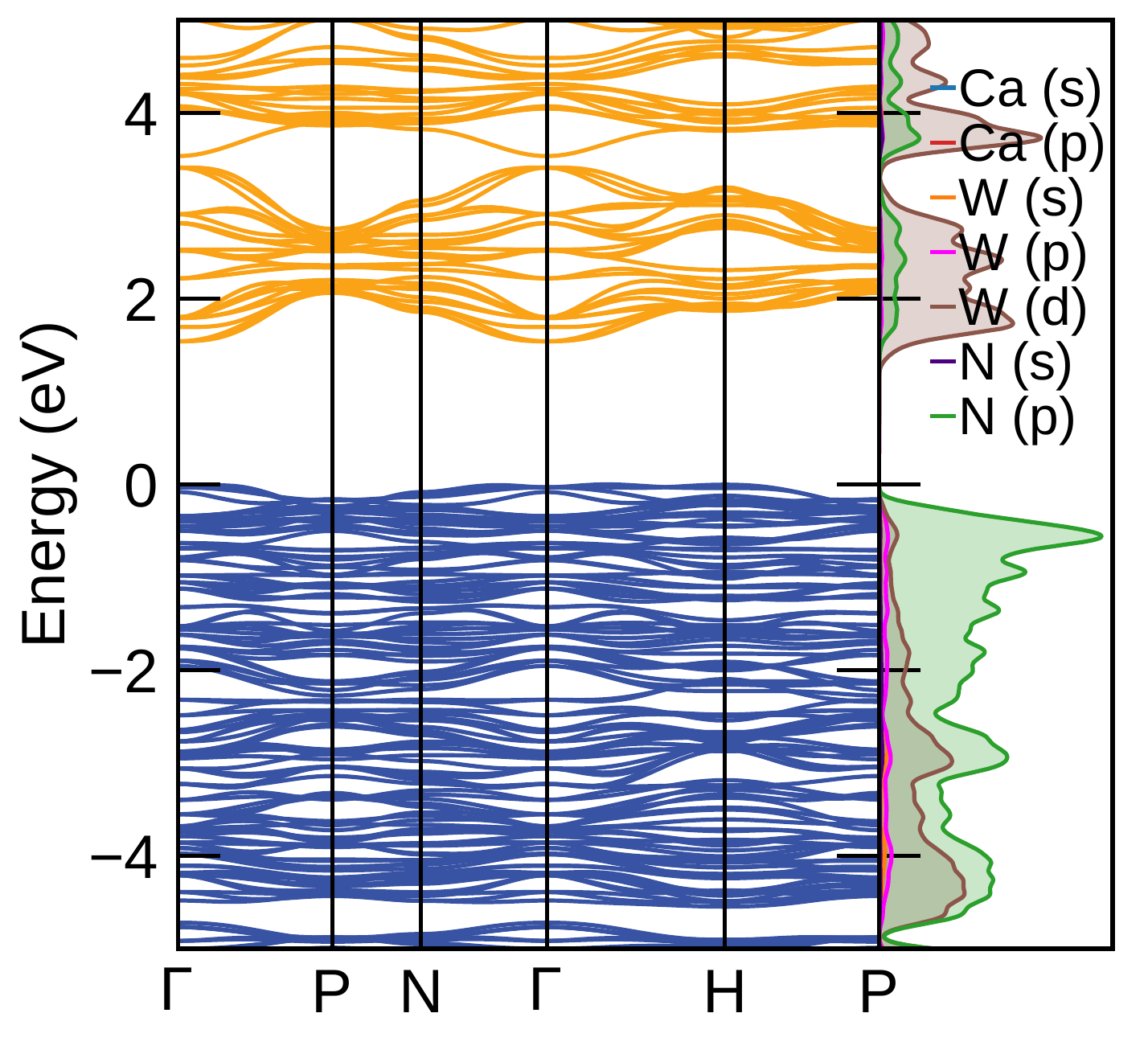} 
    \caption{\label{fig:bands_ca3w5n12}~\ce{Ca3W5N12}}
  \end{subfigure}
  \begin{subfigure}[t]{5.15cm}
    \includegraphics[width=5.15cm]{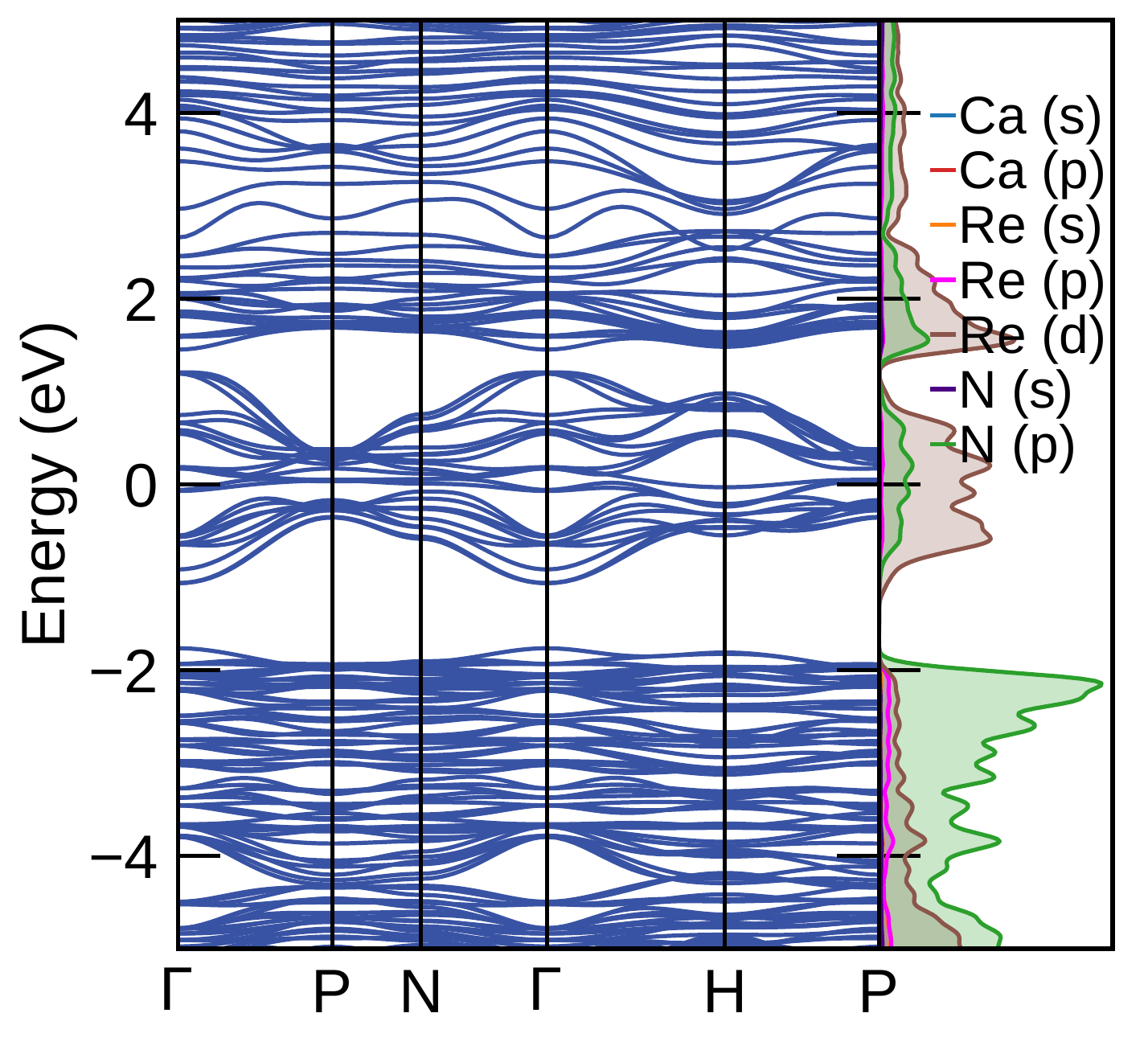}
    \caption{\label{fig:bands_ca3re5n12}~\ce{Ca3Re5N12}}
  \end{subfigure}

  \caption{Calculated mBJ electronic band structures for selected
    nitride garnets. The Fermi level is set at zero.}
  \label{fig:Band_nitrides}
\end{figure}

\begin{figure}[!th]
  \begin{subfigure}[t]{5.15cm}
    \includegraphics[width=5.15cm]{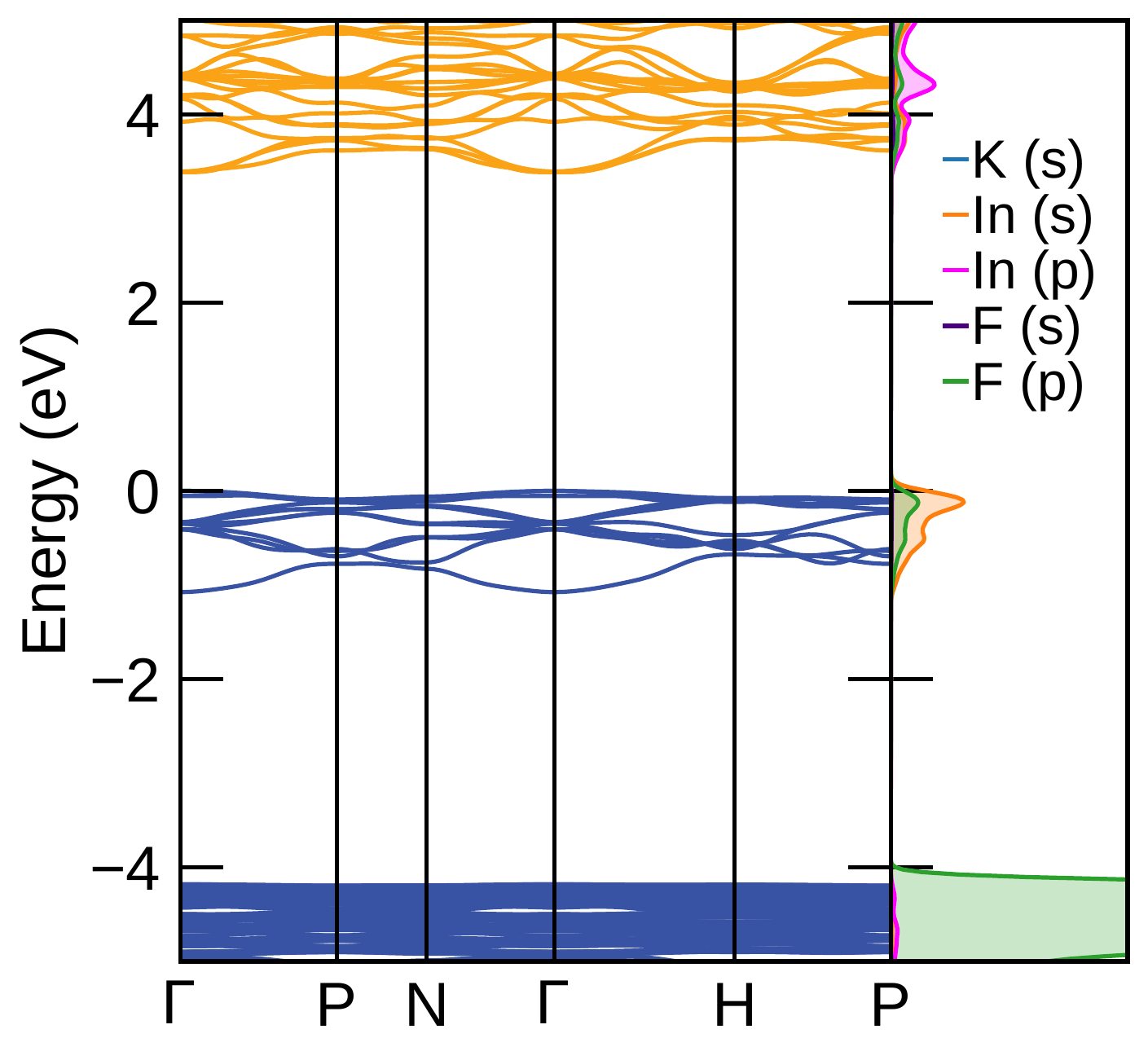} 
    \caption{\label{fig:bands_k3in5f12}~\ce{K3In5F12}}
  \end{subfigure}
  \begin{subfigure}[t]{5.15cm}
    \includegraphics[width=5.15cm]{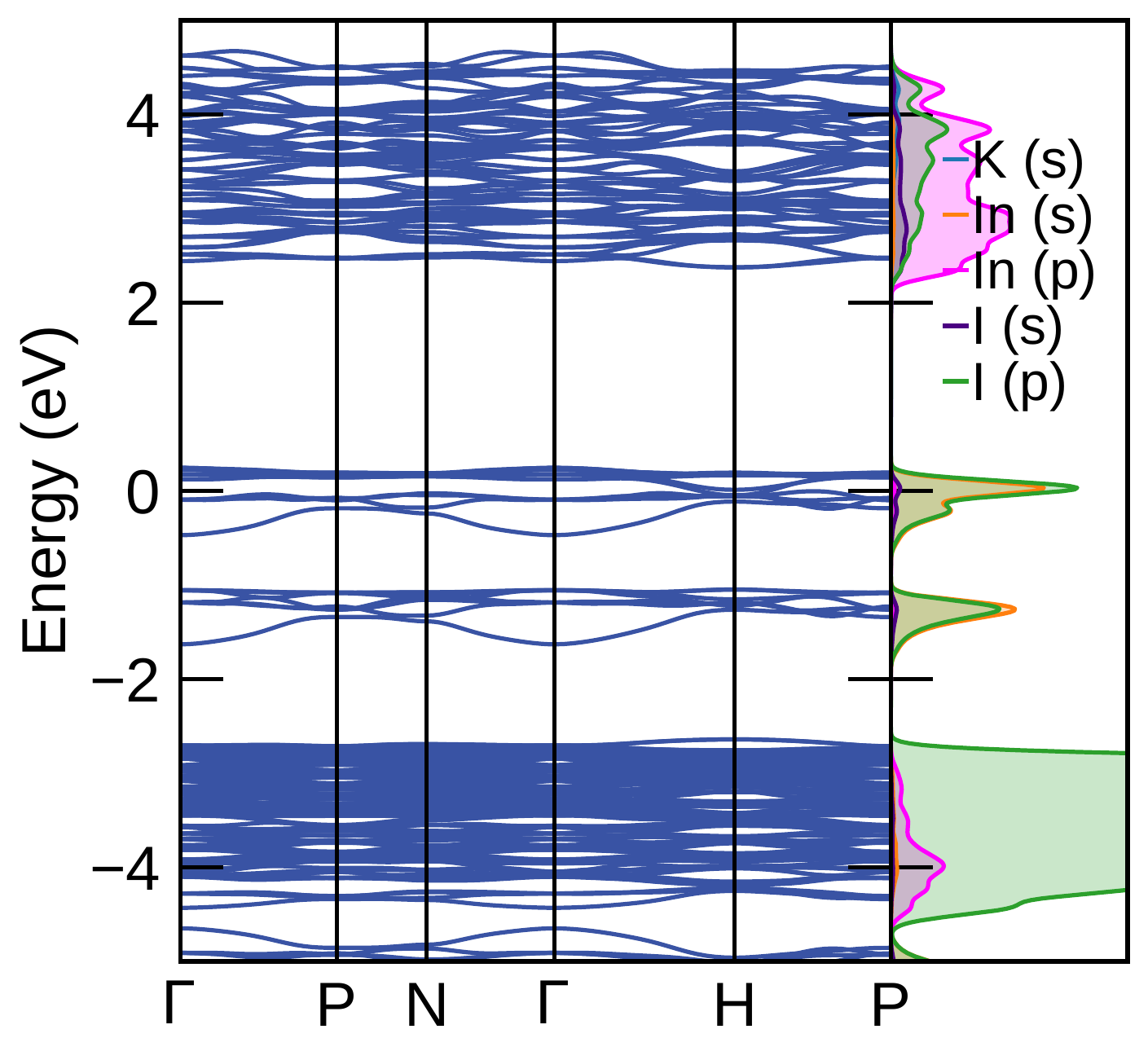}
    \caption{\label{fig:bands_k3in5i12}~\ce{K3In5I12}}
  \end{subfigure}
  \begin{subfigure}[t]{5.15cm}
    \includegraphics[width=5.15cm]{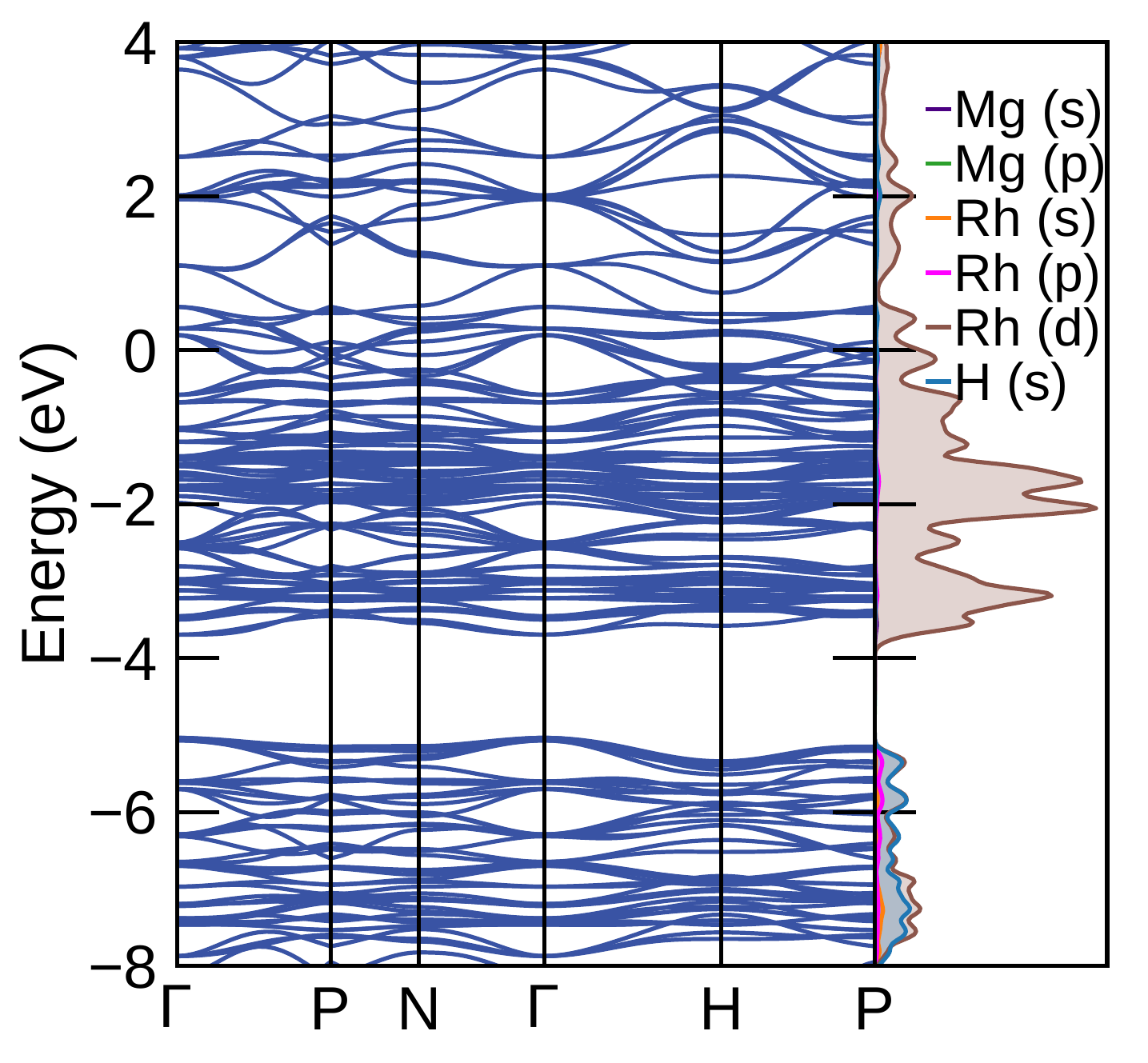} 
    \caption{\label{fig:bands_mg3rh5h12}~\ce{Mg3Rh5H12}}
  \end{subfigure}
  \begin{subfigure}[t]{5.15cm}
    \includegraphics[width=5.15cm]{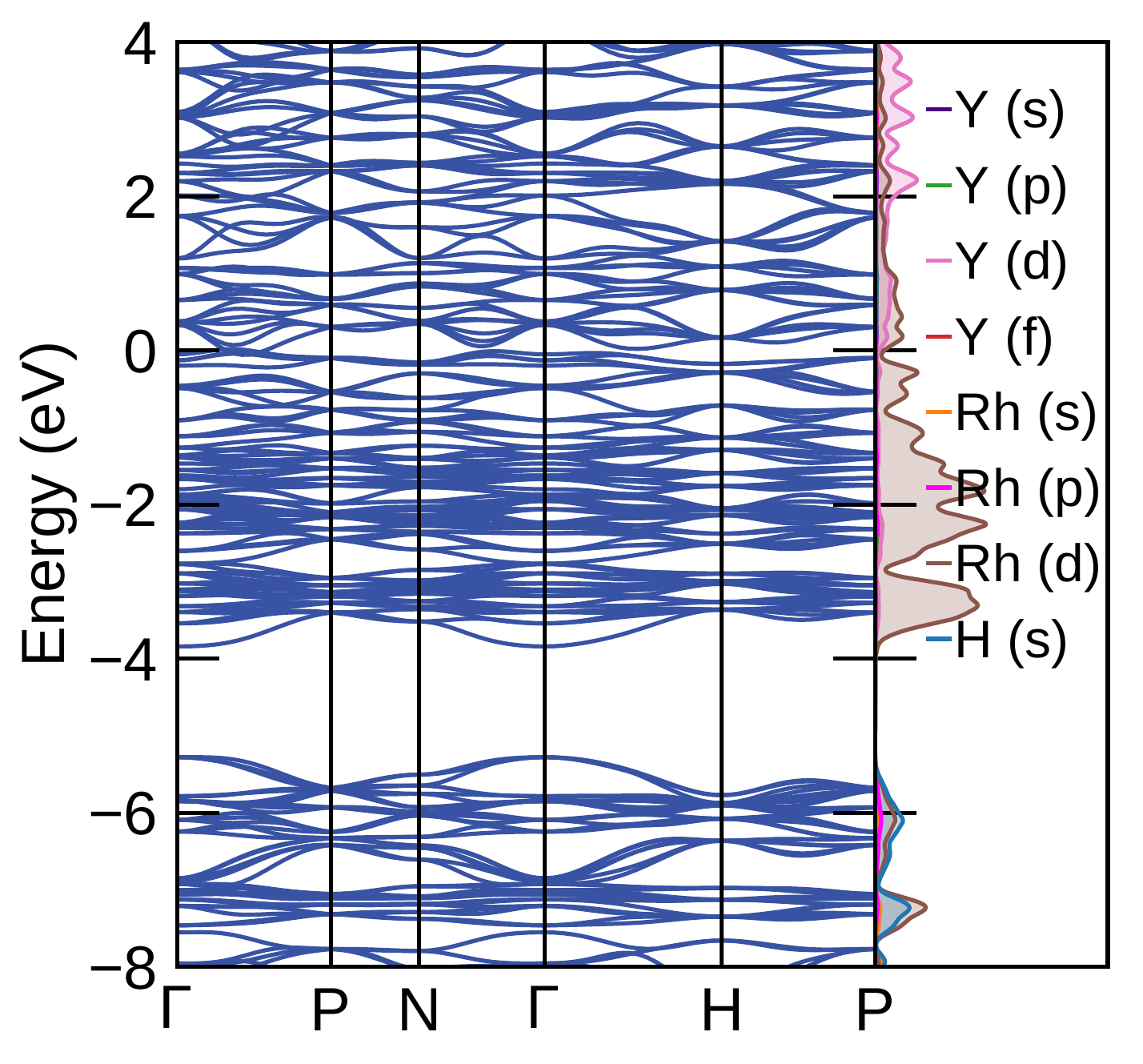}
    \caption{\label{fig:bands_y3rh5h12}~\ce{Y3Rh5H12}}
  \end{subfigure}
    
  \caption{Calculated mBJ electronic band structures for selected
    halide and hydride garnets. The Fermi level is set at zero.}
  \label{fig:Bands_halides_hydrides}
\end{figure}

We illustrate the possible electronic structures of our garnets
through a few selected examples depicted in
Figs.~\ref{fig:Band_chalcogenides}--\ref{fig:Bands_halides_hydrides}.

Generally the electron states around the Fermi level in garnets can be
classified in three categories: (i)~from s--p, p--p, or even d--p
bonding orbitals between B and C atoms, (ii)~from corresponding
anti-bonding between B and C atoms, and, (iii)~from d-electrons from A
atoms if these are d-block or f-block metals.  From the simplest
tight-binding model, and as expected from $k \cdot p$ theory, we know
that the larger the difference between the electronegativity of B and
C and the shorter B--C bond-length, the larger the separations between
states (i) and (ii). For the d (or f) block elements occupying
the A-site, the position of states (iii) can be between (i) and (ii),
overlapping with, or even above the latter. In charge compensated
situations, the bands (i) are completely filled, while (ii) and (iii)
(if applicable) bands are empty, resulting in
insulating/semiconducting systems. Otherwise, depletion of bands (i)
or filling of bands (ii) or (iii) can happen, leading to metallic
systems. Although many deviations from this simple tight-binding
picture appear, we will see how these general patterns are useful to
understand the band-structures.

For chalcogenide garnets we chose as representative examples
\ce{Y3Al5S12}, \ce{Y3Ge5S12}, and \ce{Y3Ga5S12}, while more examples
can be found in the SI. The band-structures of these compounds are
shown in Fig~\ref{fig:Band_chalcogenides}, together with \ce{Y3Al5O12}
for comparison. The oxidation-state balanced \ce{Y3Al5CH12} (CH = S,
Se, Te) garnets are semiconductors as expected. For \ce{Y3Al5S12} the
band structure is shown in Fig.~\ref{fig:bands_y3al5s12}. Similar to
its oxide counterpart in Fig.~\ref{fig:bands_y3al5o12}, the Y--$d$
states dominate the conduction bands (CB) slightly hybridizing with
Al(p) and S(p)-states. These bands are from the type (iii) states
as discussed above. The valence bands (VB) around Fermi level are
mainly composed of the localized anionic $p$--states of type (i), also
representing the typical situation described above. From \ce{Y3Al5O12}
to \ce{Y3Al5Te12} (see Fig.~S1 in SI), following the decreasing trend
of electronegativity for chalcogens, the band gap shrinks and the band
edges become more dispersed.

The elements occupying the B position in charge-compensated
chalcogenides also have a crucial effect on the electronic
structure. For example, \ce{Y3B5S12} (B = Al, Ga, In, Tl, see Fig.~S1
in the SI) all have a direct gap at $\Gamma$ with the sole exception
of \ce{Y3Al5S12} which has an indirect H--$\Gamma$ gap. Moreover,
going from Al to Tl the gap decreases, while the contibution of the
s-states from the B atoms to the bottom of the conduction bands
increases, leading to more extended bands and to lower effective
electron masses. Furthermore, for \ce{Y3Tl5S12} we can see that the Y(d)
bands are above the Tl(s)-S(s) displaying more disperse anti-bonding
mixing, an resulting in very low electron effective masses
($m^*_\text{e} = 0.19$).

For chalcogenides with unbalanced oxidation states, such as
\ce{Y3Ge5S12}, similar features can also be observed (see
Fig.~\ref{fig:bands_y3ge5s12}). Between the empty Y(d)-S(p)
hybridization states of type (iii) and the full S(p) dominated type
(i) bands are the states from Ge(s) and S(p) anti-bonding of type
(ii). As discussed above they are partially occupied, so the system is
metallic. It is possible that the system might re-establish a balanced
oxidation state by creating Ge vacancies thus becoming semiconducting,
though a detailed investigation of such a possibility is beyond the
scope of the present paper.

Some representatives of the band structures for nitrides garnets are
shown in Fig.~\ref{fig:Band_nitrides}. The oxidation state compensated
\ce{Ca3W5N12} is semiconducting. The anti-bonding $p-d$ bands forming
the CB separate also into two manifolds: the lower part is mainly
constructed from the tungsten on the tetrahedral sites, and the upper
part mainly from the octahedral W. The VB are, as usual, mostly
composed of N(p) states. For charge unbalanced nitrides (such as
\ce{Ca3Re5N12}) that have more valence electrons from Re compared to
W, the Fermi level passes through the partially occupied $p-d$
anti-bonding bands and the system becomes metallic. However, the
separation between $p-d$ bonding and anti-bonding bands can still be
seen as well as the double manifolds of $p-d$ anti-bonding states.
When a chemical element with less valence electrons replaces W, for
example, in \ce{La3Nb5N12} (see Fig.~S1 in SI), the minority spin
channel of $p-d$ bonding states is partially empty, and the system
actually becomes a half-metal.

For halides, we show the band structure of \ce{K3In5F12} and
\ce{K3In5I12} as examples. Apparently, the former compound has
unbalance oxidation states, however it is semiconducting. The Bader
net charges of the In at octahedral and tetrahedral sites are +1 and
+2.4, respectively. The In in the B-site is therefore in the +1 and +3
oxidation states, reaching charge compensation as discussed above. The
top of the valence bands are mainly composed of In(s)-F(p)
anti-bonding states, which separate from the lower valance bands,
formed by localized F(p)-In(p) bonding states, by around 4~eV. Unlike
in most sulfides and nitrides, where d-states from A atoms dominate
the bottom of the conduction bands, in \ce{K3In5F12} the bottom of the
CB is mainly formed by In(p)-F(p) anti-bonding hybridized states.
This is because the K(s)-F(p) anti-bonding states have a much higher
energy than d-p anti-bonding states in those sulfides and
nitrides. When replacing F with I, the Bader net charges for the In
atoms in octahedral and tetrahedral sites are respectively +0.7 and
+1.0, showing that the +1/+3 oxidation states is not possible in
\ce{K3In5I12}, and the system becomes metallic. The top valence band formed
by In(p)-I(s) anti-bonding states separates in two manifolds. The
upper part comes from the In atoms in tetrahedral sites and the lower
belongs to the octahedral ones. This can be partially explained by the
fact that the In--I bond length is larger in the latter.

In Fig~\ref{fig:Bands_halides_hydrides} we show the band structure of
selected examples of hydride garnets. For both \ce{Mg3Rh5H12} and
\ce{Y3Rh5H12} the separation between H(s)-Rh(d) mixed bands and cation
Rh(d) dominated bands are still present, similar to the situations in
other uncompensated garnets. These two systems are also both charge
uncompensated, the ``extra'' electrons partially filling the Rh(d)
dominated bands, leading to metallic systems. Another way to
re-establish compensation might be to force more H atoms to occupy the
interstitial sites, but again, we leave exploration of this
possibility to future investigations.

\section{Conclusions}

We performed a machine-learning assisted high-throughput investigation
of ternary garnets. We concentrated in non-oxides (that have been
studied previously) and in ferromagnetic or paramagetic compounds. We
find a wealth of systems on the convex hull (i.e., thermodynamically
stable) or close to it. This includes chalcogenides (with the
stability decreasing from S to Te), nitrides, halides, hydrides,
etc. We also found other possibilities, such as ``inverted'' garnets,
but these were slighlty less stable than the conventional
phase. The materials tend to be semiconducting/insulating when the composition is charge compensated, otherwise we obtain metallic ground-states. The latter ones could be especially relevant as, to out knowledge, no garnets conducting at room temperature are know. Some of the metallic garnets even have lattice constants that are suitable to create hetero-structures with YIG.

A few chalcogenide garnets, in particular the sulfides, are
thermodynamic stable, and are straightforward generalizations of
common oxide garnets. Band-gaps are as expected considerably smaller
for the sulfides, and decrease further across the periodic group. This
opens up the possibility to engineer the band gap of garnets by
anionic alloying, from the extreme ultaviolet of the oxide phase to
UV-A regime or even into the visible. We predict several nitride
systems that have interesting electronic properties due to the
presence of transition metals in very high charge states. In view of
the recent synthesis of two exotic nitride
perovskites~\cite{10.1126/science.abm3466,10.1002/anie.202108759} that
were predicted~\cite{Sarmiento_P_rez_2015,Flores_Livas_2019} with a
method similar to the one used in this paper, we are confident that
also nitride garnets are accessible experimentally. Finally, we find a
few semiconducting halides where the chemical element occupying the
octahedral and the dodecahedral site is in two different charge
states.

Above all, we believe that our work proves that an exhaustive survey of
the ternary, and perhaps also of the quaternary, space of materials is
now accessible to high-throughput studies, even for large and complex
unit cells. This is made possible by machine learning methods, that
already achieved an outstanding maturity in the short time since their
first appearance, and that are reaching an unprecedented accuracy. We
expect these methods to further accelerate the discovery of new
materials with exceptional properties.

\section{Acknowledgements}

The authors gratefully acknowledge the Gauss Centre for Supercomputing
e.V. (\url{www.gauss-centre.eu}) for funding this project by providing
computing time on the GCS Supercomputer SuperMUC-NG at the Leibniz
Supercomputing Centre under the project pn25co.

\bibliography{bib}
\end{document}

\author{Jonathan Schmidt}
\affiliation{\halle} 
\author{Haichen Wang}
\affiliation{\halle} 
\author{Georg Schmidt}
\affiliation{\halle} 
\author{Miguel A. L. Marques} 
\email{miguel.marques@physik.uni-halle.de}
\affiliation{\halle} 

\title{Supplementary Information: Machine Learning guided high-throughput search of non-oxide garnets} 
{
\let\clearpage\relax
\maketitle
}

\setcounter{figure}{0}
\renewcommand{\figurename}{Fig.}
\renewcommand{\thefigure}{S\arabic{figure}}

\setcounter{table}{0}
\renewcommand{\tablename}{Table}
\renewcommand{\thetable}{S\arabic{table}}

\section{Supplementary Tables}
\begin{longtable}{r c c c}
    \centering
    \caption{The full list of (meta-)stable non-oxide garnets with a distance to the convex hull of stability ($E_\text{hull}$ (in meV/Atom) below 100~meV/atom, and their calculated lattice constant ($a$, in \AA) and band gap with the PBE approximation (Gap$^{\text{PBE}}$, in eV).     
    \label{supp_all}}\\
%    \begin{tabular}{r c c }

Formula & $a$ & Gap$^{\text{PBE}}$ & $E_\text{hull}$ \\
\hline
\ce{Bi3Al5S12} & 15.121 & 2.11 & 25 \\  
\ce{Sb3Al5S12} & 14.987 & 1.76 & 76 \\  
\ce{Ba3Sb5S12} & 16.850 & 0.00 & 99 \\  
\ce{Ba3Sn5S12} & 16.602 & 0.00 & 40 \\  
\ce{Ca3Sn5S12} & 15.971 & 0.00 & 72 \\  
\ce{Ce3Al5S12} & 15.117 & 0.00 & 0 \\  
\ce{Ce3Ga5S12} & 15.239 & 0.00 & 33 \\  
\ce{Ce3Ge5S12} & 15.508 & 0.00 & 53 \\  
\ce{Ce3In5S12} & 15.905 & 0.00 & 25 \\  
\ce{Dy3Al5S12} & 14.916 & 2.07 & 0 \\  
\ce{Dy3Ga5S12} & 15.057 & 1.32 & 48 \\  
\ce{Dy3Ge5S12} & 15.305 & 0.00 & 82 \\  
\ce{Dy3In5S12} & 15.692 & 1.35 & 48 \\  
\ce{Er3Al5S12} & 14.868 & 2.12 & 9 \\  
\ce{Er3Ga5S12} & 15.006 & 1.34 & 54 \\  
\ce{Er3Ge5S12} & 15.254 & 0.00 & 98 \\  
\ce{Er3In5S12} & 15.643 & 1.70 & 65 \\  
\ce{Eu3Al5S12} & 15.107 & 0.06 & 59 \\  
\ce{Eu3As5S12} & 15.688 & 0.00 & 97 \\  
\ce{Eu3Ge5S12} & 15.494 & 0.00 & 33 \\  
\ce{Eu3In5S12} & 15.919 & 0.00 & 51 \\  
\ce{Eu3Sn5S12} & 16.145 & 0.04 & 31 \\  
\ce{Eu3Ti5S12} & 15.408 & 0.00 & 92 \\  
\ce{Bi3Ga5S12} & 15.262 & 1.94 & 67 \\  
\ce{Gd3Al5S12} & 14.983 & 1.57 & 0 \\  
\ce{Gd3Ga5S12} & 15.124 & 1.24 & 32 \\  
\ce{Gd3Ge5S12} & 15.376 & 0.00 & 65 \\  
\ce{Gd3In5S12} & 15.762 & 1.21 & 32 \\  
\ce{Ho3Al5S12} & 14.891 & 2.09 & 6 \\  
\ce{Ho3Ga5S12} & 15.033 & 1.33 & 51 \\  
\ce{Ho3In5S12} & 15.667 & 1.36 & 57 \\  
\ce{Bi3In5S12} & 15.904 & 1.90 & 56 \\  
\ce{La3Al5S12} & 15.223 & 2.05 & 0 \\  
\ce{La3Ga5S12} & 15.371 & 1.24 & 36 \\  
\ce{La3Ge5S12} & 15.639 & 0.00 & 50 \\  
\ce{La3In5S12} & 16.002 & 1.49 & 3 \\  
\ce{La3Sc5S12} & 15.828 & 2.20 & 98 \\  
\ce{La3Sn5S12} & 16.282 & 0.00 & 80 \\  
\ce{Lu3Al5S12} & 14.804 & 2.15 & 27 \\  
\ce{Lu3Ga5S12} & 14.941 & 1.36 & 74 \\  
\ce{Lu3In5S12} & 15.578 & 1.37 & 94 \\  
\ce{Nd3Al5S12} & 15.120 & 1.91 & 0 \\  
\ce{Nd3Ga5S12} & 15.266 & 1.21 & 33 \\  
\ce{Nd3Ge5S12} & 15.524 & 0.00 & 58 \\  
\ce{Nd3In5S12} & 15.900 & 1.36 & 15 \\  
\ce{Nd3Sn5S12} & 16.156 & 0.00 & 96 \\  
\ce{Pr3Al5S12} & 15.167 & 1.89 & 0 \\  
\ce{Pr3Ga5S12} & 15.314 & 1.18 & 32 \\  
\ce{Pr3Ge5S12} & 15.575 & 0.00 & 56 \\  
\ce{Pr3In5S12} & 15.947 & 1.36 & 8 \\  
\ce{Pr3Sn5S12} & 16.209 & 0.00 & 89 \\  
\ce{Sc3Al5S12} & 14.631 & 1.55 & 87 \\  
\ce{Sm3Al5S12} & 15.043 & 1.96 & 0 \\  
\ce{Sm3Ga5S12} & 15.187 & 1.27 & 37 \\  
\ce{Sm3Ge5S12} & 15.442 & 0.00 & 62 \\  
\ce{Sm3In5S12} & 15.821 & 1.35 & 25 \\  
\ce{Sr3Ge5S12} & 15.600 & 0.00 & 81 \\  
\ce{Sr3Sn5S12} & 16.259 & 0.00 & 27 \\  
\ce{Sr3Ti5S12} & 15.577 & 0.00 & 95 \\  
\ce{Tb3Al5S12} & 14.942 & 2.04 & 0 \\  
\ce{Tb3Ga5S12} & 15.084 & 1.31 & 46 \\  
\ce{Tb3Ge5S12} & 15.333 & 0.00 & 77 \\  
\ce{Tb3In5S12} & 15.718 & 1.35 & 43 \\  
\ce{Tm3Al5S12} & 14.842 & 2.13 & 15 \\  
\ce{Tm3Ga5S12} & 14.980 & 1.35 & 63 \\  
\ce{Tm3In5S12} & 15.617 & 1.36 & 78 \\  
\ce{Y3Al5S12} & 14.932 & 2.08 & 2 \\  
\ce{Y3Ga5S12} & 15.073 & 1.32 & 47 \\  
\ce{Y3Ge5S12} & 15.324 & 0.00 & 82 \\  
\ce{Y3In5S12} & 15.709 & 1.35 & 47 \\  
\ce{Bi3Al5Se12} & 15.958 & 1.61 & 72 \\  
\ce{Ba3Sb5Se12} & 17.596 & 0.00 & 86 \\  
\ce{Ba3Sn5Se12} & 17.383 & 0.00 & 38 \\  
\ce{Ca3Ge5Se12} & 16.148 & 0.00 & 90 \\  
\ce{Ca3Sn5Se12} & 16.777 & 0.00 & 70 \\  
\ce{Ce3Al5Se12} & 15.944 & 0.00 & 20 \\  
\ce{Ce3Cu5Se12} & 15.530 & 0.00 & 97 \\  
\ce{Ce3Ga5Se12} & 16.055 & 0.00 & 63 \\  
\ce{Ce3Ge5Se12} & 16.283 & 0.00 & 73 \\  
\ce{Ce3In5Se12} & 16.682 & 0.00 & 52 \\  
\ce{Dy3Al5Se12} & 15.752 & 1.46 & 36 \\  
\ce{Dy3Ga5Se12} & 15.883 & 0.51 & 81 \\  
\ce{Dy3Ge5Se12} & 16.101 & 0.00 & 88 \\  
\ce{Dy3In5Se12} & 16.485 & 0.58 & 79 \\  
\ce{Er3Al5Se12} & 15.708 & 1.49 & 51 \\  
\ce{Er3Ga5Se12} & 15.837 & 0.52 & 95 \\  
\ce{Er3In5Se12} & 16.439 & 0.57 & 97 \\  
\ce{Eu3Al5Se12} & 15.957 & 0.04 & 89 \\  
\ce{Eu3As5Se12} & 16.472 & 0.00 & 81 \\  
\ce{Eu3Ge5Se12} & 16.308 & 0.00 & 38 \\  
\ce{Eu3In5Se12} & 16.714 & 0.00 & 65 \\  
\ce{Eu3Sb5Se12} & 17.112 & 0.00 & 88 \\  
\ce{Eu3Sn5Se12} & 16.923 & 0.00 & 23 \\  
\ce{Gd3Al5Se12} & 15.812 & 1.15 & 21 \\  
\ce{Gd3Ga5Se12} & 15.949 & 0.45 & 66 \\  
\ce{Gd3Ge5Se12} & 16.167 & 0.00 & 67 \\  
\ce{Gd3In5Se12} & 16.546 & 0.56 & 57 \\  
\ce{Gd3Sn5Se12} & 16.771 & 0.00 & 98 \\  
\ce{Ho3Al5Se12} & 15.730 & 1.47 & 44 \\  
\ce{Ho3Ga5Se12} & 15.860 & 0.51 & 89 \\  
\ce{Ho3Ge5Se12} & 16.077 & 0.00 & 97 \\  
\ce{Ho3In5Se12} & 16.462 & 0.58 & 89 \\  
\ce{Bi3In5Se12} & 16.697 & 1.23 & 94 \\  
\ce{La3Al5Se12} & 16.039 & 1.38 & 19 \\  
\ce{La3Ga5Se12} & 16.182 & 0.49 & 54 \\  
\ce{La3Ge5Se12} & 16.414 & 0.00 & 55 \\  
\ce{La3In5Se12} & 16.778 & 0.73 & 25 \\  
\ce{La3Sc5Se12} & 16.568 & 1.87 & 100 \\  
\ce{La3Sn5Se12} & 17.015 & 0.00 & 77 \\  
\ce{Lu3Al5Se12} & 15.652 & 1.51 & 70 \\  
\ce{Nd3Al5Se12} & 15.942 & 1.29 & 16 \\  
\ce{Nd3Ga5Se12} & 16.084 & 0.46 & 57 \\  
\ce{Nd3Ge5Se12} & 16.301 & 0.00 & 63 \\  
\ce{Nd3In5Se12} & 16.681 & 0.60 & 36 \\  
\ce{Nd3Sn5Se12} & 16.907 & 0.00 & 91 \\  
\ce{Pr3Al5Se12} & 15.987 & 1.27 & 16 \\  
\ce{Pr3Ga5Se12} & 16.126 & 0.44 & 56 \\  
\ce{Pr3Ge5Se12} & 16.352 & 0.00 & 60 \\  
\ce{Pr3In5Se12} & 16.726 & 0.61 & 34 \\  
\ce{Pr3Sn5Se12} & 16.954 & 0.00 & 86 \\  
\ce{Sm3Al5Se12} & 15.870 & 1.34 & 18 \\  
\ce{Sm3Ga5Se12} & 16.008 & 0.49 & 63 \\  
\ce{Sm3Ge5Se12} & 16.225 & 0.00 & 65 \\  
\ce{Sm3In5Se12} & 16.607 & 0.59 & 50 \\  
\ce{Sm3Sn5Se12} & 16.831 & 0.00 & 97 \\  
\ce{Pb3Sn5Se12} & 16.997 & 0.00 & 68 \\  
\ce{Sr3Ge5Se12} & 16.433 & 0.00 & 80 \\  
\ce{Sr3Sb5Se12} & 17.271 & 0.00 & 100 \\  
\ce{Sr3Sn5Se12} & 17.053 & 0.00 & 32 \\  
\ce{Tb3Al5Se12} & 15.776 & 1.43 & 29 \\  
\ce{Tb3Ga5Se12} & 15.907 & 0.50 & 75 \\  
\ce{Tb3Ge5Se12} & 16.123 & 0.00 & 81 \\  
\ce{Tb3In5Se12} & 16.509 & 0.58 & 70 \\  
\ce{Tm3Al5Se12} & 15.685 & 1.50 & 58 \\  
\ce{Y3Al5Se12} & 15.766 & 1.45 & 37 \\  
\ce{Y3Ga5Se12} & 15.898 & 0.51 & 81 \\  
\ce{Y3Ge5Se12} & 16.116 & 0.00 & 89 \\  
\ce{Y3In5Se12} & 16.499 & 0.59 & 78 \\  
\ce{Ba3Sn5Te12} & 18.652 & 0.00 & 83 \\  
\ce{Dy3Al5Te12} & 17.120 & 0.58 & 86 \\  
\ce{Er3Al5Te12} & 17.080 & 0.59 & 98 \\  
\ce{Eu3Al5Te12} & 17.336 & 0.00 & 100 \\  
\ce{Eu3Ge5Te12} & 17.575 & 0.00 & 79 \\  
\ce{Eu3In5Te12} & 18.019 & 0.00 & 100 \\  
\ce{Eu3Si5Te12} & 17.267 & 0.00 & 98 \\  
\ce{Eu3Sn5Te12} & 18.186 & 0.00 & 84 \\  
\ce{Gd3Al5Te12} & 17.176 & 0.55 & 64 \\  
\ce{Gd3Ga5Te12} & 17.262 & 0.00 & 93 \\  
\ce{Gd3Ge5Te12} & 17.403 & 0.00 & 94 \\  
\ce{Gd3In5Te12} & 17.823 & 0.00 & 86 \\  
\ce{Ho3Al5Te12} & 17.098 & 0.59 & 92 \\  
\ce{La3Al5Te12} & 17.385 & 0.58 & 50 \\  
\ce{La3Ge5Te12} & 17.634 & 0.00 & 99 \\  
\ce{La3In5Te12} & 18.052 & 0.11 & 75 \\  
\ce{Nd3Al5Te12} & 17.292 & 0.50 & 58 \\  
\ce{Nd3Ga5Te12} & 17.384 & 0.00 & 98 \\  
\ce{Nd3In5Te12} & 17.955 & 0.00 & 79 \\  
\ce{Pr3Al5Te12} & 17.334 & 0.49 & 51 \\  
\ce{Pr3Ga5Te12} & 17.430 & 0.00 & 98 \\  
\ce{Pr3In5Te12} & 17.999 & 0.00 & 76 \\  
\ce{Sm3Al5Te12} & 17.226 & 0.54 & 64 \\  
\ce{Sm3Ga5Te12} & 17.313 & 0.00 & 98 \\  
\ce{Sm3In5Te12} & 17.886 & 0.00 & 86 \\  
\ce{Sr3Sn5Te12} & 18.345 & 0.00 & 98 \\  
\ce{Tb3Al5Te12} & 17.141 & 0.58 & 80 \\  
\ce{Y3Al5Te12} & 17.131 & 0.60 & 85 \\  
\ce{Ba3Re5N12} & 13.414 & 0.00 & 59 \\  
\ce{Ba3W5N12} & 13.485 & 0.56 & 45 \\  
\ce{Ca3Re5N12} & 12.896 & 0.00 & 9 \\  
\ce{Ca3W5N12} & 12.928 & 0.93 & 0 \\  
\ce{Ce3Mo5N12} & 12.882 & 0.03 & 16 \\  
\ce{Ce3Nb5N12} & 13.063 & 0.00 & 14 \\  
\ce{Ce3Os5N12} & 12.800 & 0.00 & 92 \\  
\ce{Ce3Re5N12} & 12.878 & 0.00 & 0 \\  
\ce{Ce3Ta5N12} & 13.051 & 0.00 & 0 \\  
\ce{Ce3W5N12} & 12.943 & 0.00 & 49 \\  
\ce{Dy3Mo5N12} & 12.766 & 0.00 & 57 \\  
\ce{Dy3Re5N12} & 12.746 & 0.00 & 0 \\  
\ce{Dy3W5N12} & 12.799 & 0.00 & 72 \\  
\ce{Er3Mo5N12} & 12.710 & 0.00 & 69 \\  
\ce{Er3Re5N12} & 12.693 & 0.00 & 0 \\  
\ce{Er3W5N12} & 12.745 & 0.00 & 85 \\  
\ce{Eu3Mo5N12} & 12.990 & 0.00 & 0 \\  
\ce{Eu3Os5N12} & 12.910 & 0.00 & 75 \\  
\ce{Eu3Re5N12} & 12.985 & 0.00 & 0 \\  
\ce{Eu3W5N12} & 13.038 & 0.00 & 0 \\  
\ce{Gd3Mo5N12} & 12.851 & 0.00 & 44 \\  
\ce{Gd3Nb5N12} & 13.037 & 0.00 & 70 \\  
\ce{Gd3Os5N12} & 12.776 & 0.00 & 100 \\  
\ce{Gd3Re5N12} & 12.829 & 0.00 & 0 \\  
\ce{Gd3Ta5N12} & 13.015 & 0.00 & 70 \\  
\ce{Gd3W5N12} & 12.888 & 0.00 & 56 \\  
\ce{Ho3Mo5N12} & 12.739 & 0.00 & 63 \\  
\ce{Ho3Re5N12} & 12.720 & 0.00 & 0 \\  
\ce{Ho3W5N12} & 12.772 & 0.00 & 79 \\  
\ce{La3Mo5N12} & 13.108 & 0.00 & 44 \\  
\ce{La3Nb5N12} & 13.297 & 0.00 & 100 \\  
\ce{La3Os5N12} & 13.008 & 0.00 & 42 \\  
\ce{La3Re5N12} & 13.069 & 0.00 & 0 \\  
\ce{La3Ta5N12} & 13.276 & 0.00 & 85 \\  
\ce{La3W5N12} & 13.141 & 0.00 & 51 \\  
\ce{Lu3Re5N12} & 12.618 & 0.00 & 0 \\  
\ce{Nd3Mo5N12} & 12.990 & 0.00 & 24 \\  
\ce{Nd3Os5N12} & 12.902 & 0.00 & 48 \\  
\ce{Nd3Re5N12} & 12.956 & 0.00 & 0 \\  
\ce{Nd3Ta5N12} & 13.169 & 0.00 & 81 \\  
\ce{Nd3W5N12} & 13.020 & 0.00 & 40 \\  
\ce{Pr3Mo5N12} & 13.040 & 0.00 & 21 \\  
\ce{Pr3Os5N12} & 12.947 & 0.00 & 37 \\  
\ce{Pr3Re5N12} & 13.002 & 0.00 & 0 \\  
\ce{Pr3Ta5N12} & 13.218 & 0.00 & 81 \\  
\ce{Pr3W5N12} & 13.068 & 0.00 & 41 \\  
\ce{Pb3Re5N12} & 13.202 & 0.00 & 76 \\  
\ce{Sm3Mo5N12} & 12.908 & 0.00 & 33 \\  
\ce{Sm3Os5N12} & 12.826 & 0.00 & 77 \\  
\ce{Sm3Re5N12} & 12.880 & 0.00 & 0 \\  
\ce{Sm3Ta5N12} & 13.087 & 0.00 & 96 \\  
\ce{Sm3W5N12} & 12.939 & 0.00 & 46 \\  
\ce{Sr3Mo5N12} & 13.158 & 0.32 & 98 \\  
\ce{Sr3Re5N12} & 13.126 & 0.00 & 8 \\  
\ce{Sr3W5N12} & 13.174 & 0.89 & 0 \\  
\ce{Tb3Mo5N12} & 12.795 & 0.00 & 50 \\  
\ce{Tb3Re5N12} & 12.773 & 0.00 & 0 \\  
\ce{Tb3W5N12} & 12.828 & 0.00 & 65 \\  
\ce{Tm3Mo5N12} & 12.678 & 0.00 & 77 \\  
\ce{Tm3Re5N12} & 12.661 & 0.00 & 0 \\  
\ce{Tm3W5N12} & 12.713 & 0.00 & 93 \\  
\ce{Y3Mo5N12} & 12.787 & 0.00 & 68 \\  
\ce{Y3Re5N12} & 12.767 & 0.00 & 0 \\  
\ce{Y3W5N12} & 12.820 & 0.00 & 73 \\  
\ce{Ag3Hg5F12} & 14.122 & 0.00 & 75 \\  
\ce{Hg3Ag5F12} & 13.814 & 0.04 & 70 \\  
\ce{Pb3Ag5F12} & 14.198 & 0.00 & 45 \\  
\ce{Ba3Ag5F12} & 14.580 & 0.00 & 41 \\  
\ce{Ba3Li5F12} & 13.385 & 0.00 & 80 \\  
\ce{Ba3Na5F12} & 14.281 & 0.00 & 64 \\  
\ce{Cs3Ag5F12} & 14.924 & 0.00 & 92 \\  
\ce{Cs3Cd5F12} & 15.055 & 0.00 & 98 \\  
\ce{Cs3Cu5F12} & 13.942 & 0.00 & 92 \\  
\ce{Cs3Hg5F12} & 15.331 & 0.00 & 71 \\  
\ce{Cs3In5F12} & 15.557 & 2.82 & 27 \\  
\ce{Cs3Tl5F12} & 15.831 & 2.51 & 14 \\  
\ce{Hg3Cs5F12} & 16.374 & 0.00 & 45 \\  
\ce{Ag3Cu5F12} & 13.133 & 0.00 & 53 \\  
\ce{Eu3Ag5F12} & 14.052 & 0.00 & 64 \\  
\ce{Eu3Tl5F12} & 14.945 & 0.00 & 84 \\  
\ce{K3Ag5F12} & 14.354 & 0.03 & 71 \\  
\ce{K3Cd5F12} & 14.437 & 0.00 & 75 \\  
\ce{K3Cu5F12} & 13.389 & 0.00 & 51 \\  
\ce{K3Hg5F12} & 14.694 & 0.00 & 64 \\  
\ce{K3In5F12} & 14.898 & 2.41 & 59 \\  
\ce{K3Tl5F12} & 15.157 & 2.28 & 29 \\  
\ce{K3Zn5F12} & 13.592 & 0.00 & 84 \\  
\ce{Ba3K5F12} & 15.276 & 0.00 & 92 \\  
\ce{Eu3K5F12} & 14.547 & 0.00 & 68 \\  
\ce{Hg3K5F12} & 15.221 & 0.00 & 55 \\  
\ce{Li3Cu5F12} & 12.795 & 0.00 & 89 \\  
\ce{Ca3Li5F12} & 12.444 & 0.00 & 47 \\  
\ce{Cd3Li5F12} & 12.362 & 0.00 & 33 \\  
\ce{Ce3Li5F12} & 12.794 & 0.00 & 76 \\  
\ce{Eu3Li5F12} & 12.742 & 0.00 & 0 \\  
\ce{Gd3Li5F12} & 12.565 & 0.00 & 86 \\  
\ce{Hg3Li5F12} & 12.583 & 0.00 & 19 \\  
\ce{In3Li5F12} & 12.434 & 0.00 & 100 \\  
\ce{La3Li5F12} & 12.793 & 0.00 & 87 \\  
\ce{Pb3Li5F12} & 12.883 & 0.00 & 52 \\  
\ce{Tl3Li5F12} & 12.662 & 0.00 & 62 \\  
\ce{Na3Ag5F12} & 14.202 & 0.00 & 78 \\  
\ce{Na3Cu5F12} & 12.879 & 0.00 & 34 \\  
\ce{Na3Hg5F12} & 14.074 & 0.00 & 99 \\  
\ce{Na3Zn5F12} & 13.040 & 0.00 & 75 \\  
\ce{Eu3Na5F12} & 13.575 & 0.00 & 0 \\  
\ce{Hg3Na5F12} & 13.375 & 0.00 & 87 \\  
\ce{Pb3Na5F12} & 13.723 & 0.00 & 74 \\  
\ce{Sr3Na5F12} & 13.788 & 0.00 & 76 \\  
\ce{Rb3Ag5F12} & 14.617 & 0.01 & 84 \\  
\ce{Rb3Cd5F12} & 14.716 & 0.00 & 81 \\  
\ce{Rb3Cu5F12} & 13.644 & 0.00 & 63 \\  
\ce{Rb3Hg5F12} & 14.977 & 0.00 & 59 \\  
\ce{Rb3In5F12} & 15.204 & 2.56 & 36 \\  
\ce{Rb3Tl5F12} & 15.466 & 2.34 & 7 \\  
\ce{Rb3Zn5F12} & 13.862 & 0.00 & 98 \\  
\ce{Hg3Rb5F12} & 15.714 & 0.00 & 47 \\  
\ce{Sr3Ag5F12} & 14.113 & 0.00 & 66 \\  
\ce{Sr3Li5F12} & 12.879 & 0.00 & 58 \\  
\ce{Tl3Ag5F12} & 13.940 & 0.00 & 91 \\  
\ce{Tl3Cd5F12} & 14.674 & 0.00 & 92 \\  
\ce{Tl3Cu5F12} & 13.647 & 0.00 & 58 \\  
\ce{Tl3Hg5F12} & 14.916 & 0.00 & 85 \\  
\ce{Tl3In5F12} & 15.071 & 2.38 & 42 \\  
\ce{Tl3Zn5F12} & 13.852 & 0.00 & 96 \\  
\ce{Ag3Tl5F12} & 14.576 & 1.70 & 72 \\  
\ce{Hg3Ag5Cl12} & 16.772 & 0.00 & 84 \\  
\ce{Pb3Ag5Cl12} & 16.364 & 0.00 & 22 \\  
\ce{Sn3Ag5Cl12} & 16.252 & 0.00 & 76 \\  
\ce{Ba3Ag5Cl12} & 16.772 & 0.00 & 15 \\  
\ce{Ba3Cu5Cl12} & 15.976 & 0.00 & 51 \\  
\ce{Ba3Li5Cl12} & 16.193 & 0.00 & 74 \\  
\ce{Ba3Na5Cl12} & 17.112 & 0.00 & 92 \\  
\ce{Ca3Ag5Cl12} & 16.060 & 0.00 & 69 \\  
\ce{Ca3Cu5Cl12} & 15.271 & 0.00 & 66 \\  
\ce{In3Cd5Cl12} & 17.067 & 0.00 & 66 \\  
\ce{Cs3Ag5Cl12} & 17.198 & 0.00 & 80 \\  
\ce{Cs3Cd5Cl12} & 17.549 & 0.00 & 59 \\  
\ce{Cs3Cu5Cl12} & 16.344 & 0.00 & 89 \\  
\ce{Cs3Hg5Cl12} & 17.724 & 0.00 & 55 \\  
\ce{Cs3Sr5Cl12} & 18.711 & 0.00 & 99 \\  
\ce{Cs3Tl5Cl12} & 18.609 & 0.00 & 73 \\  
\ce{Eu3Cs5Cl12} & 20.205 & 0.00 & 63 \\  
\ce{Cu3Ag5Cl12} & 15.819 & 0.00 & 42 \\  
\ce{Ag3Cu5Cl12} & 15.337 & 0.00 & 49 \\  
\ce{Hg3Cu5Cl12} & 15.452 & 0.03 & 98 \\  
\ce{Pb3Cu5Cl12} & 15.598 & 0.00 & 35 \\  
\ce{Sn3Cu5Cl12} & 15.484 & 0.00 & 47 \\  
\ce{Eu3Ag5Cl12} & 16.255 & 0.00 & 0 \\  
\ce{Eu3Cu5Cl12} & 15.457 & 0.00 & 17 \\  
\ce{Hg3Ga5Cl12} & 17.510 & 0.00 & 97 \\  
\ce{K3Ag5Cl12} & 16.656 & 0.00 & 43 \\  
\ce{K3Cd5Cl12} & 16.983 & 0.00 & 51 \\  
\ce{K3Cu5Cl12} & 15.783 & 0.00 & 25 \\  
\ce{K3Hg5Cl12} & 17.145 & 0.00 & 39 \\  
\ce{K3Mg5Cl12} & 16.485 & 0.00 & 93 \\  
\ce{K3Tl5Cl12} & 17.993 & 0.00 & 80 \\  
\ce{K3Zn5Cl12} & 16.224 & 0.00 & 92 \\  
\ce{Li3Ag5Cl12} & 16.198 & 0.00 & 73 \\  
\ce{Li3Cu5Cl12} & 15.192 & 0.00 & 83 \\  
\ce{Ca3Li5Cl12} & 15.390 & 0.00 & 91 \\  
\ce{Ce3Li5Cl12} & 15.549 & 0.00 & 25 \\  
\ce{Eu3Li5Cl12} & 15.530 & 0.00 & 0 \\  
\ce{Gd3Li5Cl12} & 15.343 & 0.00 & 44 \\  
\ce{La3Li5Cl12} & 15.546 & 0.00 & 47 \\  
\ce{Nd3Li5Cl12} & 15.357 & 0.00 & 69 \\  
\ce{Pb3Li5Cl12} & 15.651 & 0.00 & 53 \\  
\ce{Pr3Li5Cl12} & 15.414 & 0.00 & 62 \\  
\ce{Sm3Li5Cl12} & 15.262 & 0.00 & 83 \\  
\ce{Sn3Li5Cl12} & 15.507 & 0.00 & 91 \\  
\ce{Na3Ag5Cl12} & 16.278 & 0.00 & 42 \\  
\ce{Na3Cd5Cl12} & 16.520 & 0.00 & 86 \\  
\ce{Na3Cu5Cl12} & 15.343 & 0.00 & 27 \\  
\ce{Na3Hg5Cl12} & 16.702 & 0.00 & 90 \\  
\ce{Ce3Na5Cl12} & 16.438 & 0.00 & 92 \\  
\ce{Eu3Na5Cl12} & 16.373 & 0.00 & 43 \\  
\ce{Hg3Na5Cl12} & 16.891 & 0.00 & 96 \\  
\ce{Rb3Ag5Cl12} & 16.899 & 0.00 & 59 \\  
\ce{Rb3Ca5Cl12} & 17.797 & 0.00 & 98 \\  
\ce{Rb3Cd5Cl12} & 17.229 & 0.00 & 47 \\  
\ce{Rb3Cu5Cl12} & 16.027 & 0.00 & 51 \\  
\ce{Rb3Eu5Cl12} & 18.271 & 0.00 & 93 \\  
\ce{Rb3Hg5Cl12} & 17.400 & 0.00 & 41 \\  
\ce{Rb3Sr5Cl12} & 18.390 & 0.00 & 97 \\  
\ce{Rb3Tl5Cl12} & 18.260 & 0.00 & 72 \\  
\ce{Eu3Rb5Cl12} & 19.508 & 0.00 & 75 \\  
\ce{Hg3Rb5Cl12} & 18.448 & 0.00 & 60 \\  
\ce{Sr3Ag5Cl12} & 16.384 & 0.00 & 25 \\  
\ce{Sr3Cu5Cl12} & 15.585 & 0.00 & 41 \\  
\ce{Sr3Li5Cl12} & 15.757 & 0.00 & 66 \\  
\ce{Tl3Ag5Cl12} & 16.722 & 0.00 & 64 \\  
\ce{Tl3Cd5Cl12} & 17.098 & 0.00 & 59 \\  
\ce{Tl3Cu5Cl12} & 15.905 & 0.00 & 34 \\  
\ce{Tl3Hg5Cl12} & 17.262 & 0.00 & 55 \\  
\ce{Au3Ag5Br12} & 17.007 & 0.00 & 86 \\  
\ce{Hg3Ag5Br12} & 17.623 & 0.00 & 93 \\  
\ce{Pb3Ag5Br12} & 17.162 & 0.00 & 40 \\  
\ce{Sn3Ag5Br12} & 17.060 & 0.00 & 81 \\  
\ce{Ba3Ag5Br12} & 17.602 & 0.00 & 19 \\  
\ce{Ba3Cu5Br12} & 16.847 & 0.00 & 53 \\  
\ce{Ba3Li5Br12} & 17.125 & 0.00 & 66 \\  
\ce{Ba3Na5Br12} & 18.053 & 0.00 & 88 \\  
\ce{Ba3Tl5Br12} & 18.995 & 0.00 & 98 \\  
\ce{Ca3Ag5Br12} & 16.923 & 0.00 & 82 \\  
\ce{Ca3Cu5Br12} & 16.186 & 0.00 & 87 \\  
\ce{Ga3Cd5Br12} & 17.556 & 0.00 & 94 \\  
\ce{In3Cd5Br12} & 17.868 & 0.00 & 59 \\  
\ce{Cs3Ag5Br12} & 18.073 & 0.00 & 74 \\  
\ce{Cs3Ca5Br12} & 18.984 & 0.00 & 99 \\  
\ce{Cs3Cd5Br12} & 18.416 & 0.00 & 48 \\  
\ce{Cs3Cu5Br12} & 17.262 & 0.00 & 86 \\  
\ce{Cs3Hg5Br12} & 18.583 & 0.00 & 44 \\  
\ce{Cs3In5Br12} & 19.273 & 0.00 & 75 \\  
\ce{Cs3Pb5Br12} & 19.488 & 0.00 & 100 \\  
\ce{Cs3Sr5Br12} & 19.585 & 0.00 & 88 \\  
\ce{Cs3Tl5Br12} & 19.458 & 0.00 & 49 \\  
\ce{Ag3Cu5Br12} & 16.313 & 0.00 & 67 \\  
\ce{Pb3Cu5Br12} & 16.444 & 0.00 & 52 \\  
\ce{Sn3Cu5Br12} & 16.340 & 0.00 & 63 \\  
\ce{Eu3Ag5Br12} & 17.100 & 0.00 & 2 \\  
\ce{Eu3Cu5Br12} & 16.356 & 0.00 & 36 \\  
\ce{Hg3Ga5Br12} & 18.161 & 0.00 & 95 \\  
\ce{Hg3In5Br12} & 18.832 & 0.00 & 94 \\  
\ce{K3Ag5Br12} & 17.508 & 0.00 & 44 \\  
\ce{K3Cd5Br12} & 17.880 & 0.00 & 34 \\  
\ce{K3Cu5Br12} & 16.714 & 0.00 & 29 \\  
\ce{K3Eu5Br12} & 18.907 & 0.00 & 97 \\  
\ce{K3Hg5Br12} & 18.012 & 0.00 & 29 \\  
\ce{K3In5Br12} & 18.685 & 0.00 & 87 \\  
\ce{K3Mg5Br12} & 17.489 & 0.00 & 84 \\  
\ce{K3Sr5Br12} & 19.066 & 0.00 & 95 \\  
\ce{K3Tl5Br12} & 18.849 & 0.00 & 65 \\  
\ce{K3Zn5Br12} & 17.154 & 0.00 & 95 \\  
\ce{Li3Ag5Br12} & 17.136 & 0.00 & 80 \\  
\ce{Li3Cu5Br12} & 16.204 & 0.00 & 86 \\  
\ce{Ce3Li5Br12} & 16.505 & 0.00 & 57 \\  
\ce{Eu3Li5Br12} & 16.495 & 0.00 & 0 \\  
\ce{Gd3Li5Br12} & 16.314 & 0.00 & 77 \\  
\ce{La3Li5Br12} & 16.495 & 0.00 & 68 \\  
\ce{Nd3Li5Br12} & 16.328 & 0.00 & 92 \\  
\ce{Pb3Li5Br12} & 16.620 & 0.00 & 76 \\  
\ce{Pr3Li5Br12} & 16.377 & 0.00 & 87 \\  
\ce{Na3Ag5Br12} & 17.233 & 0.00 & 58 \\  
\ce{Na3Cd5Br12} & 17.455 & 0.00 & 82 \\  
\ce{Na3Cu5Br12} & 16.293 & 0.00 & 40 \\  
\ce{Na3Hg5Br12} & 17.581 & 0.00 & 76 \\  
\ce{Cu3Na5Br12} & 17.052 & 0.00 & 64 \\  
\ce{Eu3Na5Br12} & 17.343 & 0.00 & 33 \\  
\ce{Rb3Ag5Br12} & 17.774 & 0.00 & 51 \\  
\ce{Rb3Ca5Br12} & 18.702 & 0.00 & 94 \\  
\ce{Rb3Cd5Br12} & 18.116 & 0.00 & 36 \\  
\ce{Rb3Cu5Br12} & 16.951 & 0.00 & 50 \\  
\ce{Rb3Hg5Br12} & 18.278 & 0.00 & 31 \\  
\ce{Rb3In5Br12} & 18.957 & 0.00 & 79 \\  
\ce{Rb3Mg5Br12} & 17.723 & 0.00 & 90 \\  
\ce{Rb3Sr5Br12} & 19.271 & 0.00 & 88 \\  
\ce{Rb3Tl5Br12} & 19.141 & 0.00 & 49 \\  
\ce{Eu3Rb5Br12} & 20.331 & 0.00 & 78 \\  
\ce{Sr3Ag5Br12} & 17.238 & 0.00 & 31 \\  
\ce{Sr3Cu5Br12} & 16.486 & 0.00 & 50 \\  
\ce{Sr3Li5Br12} & 16.721 & 0.00 & 67 \\  
\ce{Tl3Ag5Br12} & 17.530 & 0.00 & 63 \\  
\ce{Tl3Cd5Br12} & 17.907 & 0.00 & 48 \\  
\ce{Tl3Cu5Br12} & 16.764 & 0.00 & 43 \\  
\ce{Tl3Hg5Br12} & 18.058 & 0.00 & 41 \\  
\ce{Tl3In5Br12} & 18.639 & 0.00 & 77 \\  
\ce{Pb3Ag5I12} & 18.251 & 0.00 & 88 \\  
\ce{Ba3Ag5I12} & 18.731 & 0.00 & 36 \\  
\ce{Ba3Au5I12} & 18.638 & 0.00 & 99 \\  
\ce{Ba3Cu5I12} & 18.028 & 0.00 & 45 \\  
\ce{Ba3Ga5I12} & 19.525 & 0.00 & 93 \\  
\ce{Ba3In5I12} & 20.102 & 0.00 & 98 \\  
\ce{Ba3Li5I12} & 18.492 & 0.00 & 55 \\  
\ce{Ba3Na5I12} & 19.424 & 0.00 & 79 \\  
\ce{Ba3Tl5I12} & 20.247 & 0.00 & 91 \\  
\ce{In3Cd5I12} & 19.029 & 0.00 & 66 \\  
\ce{Cs3Cd5I12} & 19.701 & 0.00 & 55 \\  
\ce{Cs3Cu5I12} & 18.488 & 0.00 & 97 \\  
\ce{Cs3Ga5I12} & 19.797 & 0.00 & 72 \\  
\ce{Cs3Hg5I12} & 19.803 & 0.00 & 54 \\  
\ce{Cs3In5I12} & 20.504 & 0.00 & 53 \\  
\ce{Cs3Mg5I12} & 19.470 & 0.00 & 89 \\  
\ce{Cs3Pb5I12} & 20.741 & 0.00 & 100 \\  
\ce{Cs3Sr5I12} & 20.999 & 0.00 & 98 \\  
\ce{Cs3Tl5I12} & 20.690 & 0.00 & 66 \\  
\ce{Pb3Cu5I12} & 17.581 & 0.00 & 74 \\  
\ce{Sn3Cu5I12} & 17.465 & 0.00 & 90 \\  
\ce{Eu3Ag5I12} & 18.253 & 0.00 & 52 \\  
\ce{Eu3Cu5I12} & 17.552 & 0.00 & 53 \\  
\ce{Hg3Ga5I12} & 19.164 & 0.00 & 88 \\  
\ce{In3Hg5I12} & 19.096 & 0.00 & 88 \\  
\ce{Hg3In5I12} & 19.857 & 0.00 & 75 \\  
\ce{K3Ag5I12} & 18.703 & 0.00 & 82 \\  
\ce{K3Au5I12} & 18.647 & 0.00 & 89 \\  
\ce{K3Cd5I12} & 19.204 & 0.00 & 48 \\  
\ce{K3Cu5I12} & 17.963 & 0.00 & 46 \\  
\ce{K3Ga5I12} & 19.305 & 0.00 & 99 \\  
\ce{K3Hg5I12} & 19.285 & 0.00 & 45 \\  
\ce{K3In5I12} & 19.977 & 0.00 & 61 \\  
\ce{K3Mg5I12} & 18.946 & 0.00 & 86 \\  
\ce{K3Tl5I12} & 20.120 & 0.00 & 77 \\  
\ce{Hg3K5I12} & 20.209 & 0.00 & 94 \\  
\ce{Ce3Li5I12} & 17.882 & 0.00 & 75 \\  
\ce{Eu3Li5I12} & 17.909 & 0.00 & 35 \\  
\ce{La3Li5I12} & 17.857 & 0.00 & 99 \\  
\ce{Na3Ag5I12} & 18.672 & 0.00 & 99 \\  
\ce{Na3Au5I12} & 18.456 & 0.00 & 91 \\  
\ce{Na3Cd5I12} & 18.788 & 0.00 & 90 \\  
\ce{Na3Cu5I12} & 17.579 & 0.00 & 53 \\  
\ce{Na3Hg5I12} & 18.872 & 0.00 & 86 \\  
\ce{Eu3Na5I12} & 18.783 & 0.00 & 84 \\  
\ce{Rb3Ag5I12} & 18.945 & 0.00 & 90 \\  
\ce{Rb3Au5I12} & 18.873 & 0.00 & 100 \\  
\ce{Rb3Cd5I12} & 19.427 & 0.00 & 47 \\  
\ce{Rb3Cu5I12} & 18.192 & 0.00 & 63 \\  
\ce{Rb3Ga5I12} & 19.504 & 0.00 & 69 \\  
\ce{Rb3Hg5I12} & 19.524 & 0.00 & 41 \\  
\ce{Rb3In5I12} & 20.218 & 0.00 & 63 \\  
\ce{Rb3Mg5I12} & 19.179 & 0.00 & 82 \\  
\ce{Rb3Sr5I12} & 20.716 & 0.00 & 100 \\  
\ce{Rb3Tl5I12} & 20.389 & 0.00 & 70 \\  
\ce{Hg3Rb5I12} & 20.730 & 0.00 & 98 \\  
\ce{Sr3Ag5I12} & 18.405 & 0.00 & 63 \\  
\ce{Sr3Cu5I12} & 17.701 & 0.00 & 59 \\  
\ce{Sr3Li5I12} & 18.132 & 0.00 & 78 \\  
\ce{Tl3Ag5I12} & 18.613 & 0.00 & 89 \\  
\ce{Tl3Au5I12} & 18.550 & 0.00 & 77 \\  
\ce{Tl3Cd5I12} & 19.089 & 0.00 & 59 \\  
\ce{Tl3Cu5I12} & 17.916 & 0.00 & 50 \\  
\ce{Tl3Ga5I12} & 19.117 & 0.00 & 82 \\  
\ce{Tl3Hg5I12} & 19.173 & 0.00 & 55 \\  
\ce{Tl3In5I12} & 19.793 & 0.00 & 72 \\  
\ce{Ag3Tl5I12} & 19.569 & 0.00 & 83 \\  
\ce{Au3Tl5I12} & 19.583 & 0.00 & 48 \\  
\ce{Ag3Pt5H12} & 11.540 & 0.00 & 90 \\  
\ce{Ca3Cu5H12} & 11.635 & 0.00 & 87 \\  
\ce{Cu3Rh5H12} & 10.686 & 0.00 & 96 \\  
\ce{Dy3Rh5H12} & 11.560 & 0.26 & 95 \\  
\ce{Er3Rh5H12} & 11.490 & 0.25 & 95 \\  
\ce{Gd3Pd5H12} & 11.937 & 0.00 & 98 \\  
\ce{Gd3Rh5H12} & 11.645 & 0.06 & 90 \\  
\ce{Pd3Pt5H12} & 11.154 & 0.00 & 72 \\  
\ce{Pt3Pd5H12} & 11.091 & 0.00 & 100 \\  
\ce{Rh3Pd5H12} & 10.879 & 0.00 & 79 \\  
\ce{Pb3Rh5H12} & 11.850 & 0.15 & 96 \\  
\ce{Hf3Mo5H12} & 11.195 & 0.00 & 85 \\  
\ce{Hf3Nb5H12} & 11.455 & 0.00 & 39 \\  
\ce{Hf3Ta5H12} & 11.394 & 0.00 & 32 \\  
\ce{Hf3Ti5H12} & 11.277 & 0.00 & 43 \\  
\ce{Hf3Zr5H12} & 11.884 & 0.00 & 61 \\  
\ce{Ti3Hf5H12} & 11.480 & 0.00 & 81 \\  
\ce{Zr3Hf5H12} & 11.844 & 0.00 & 57 \\  
\ce{Ho3Rh5H12} & 11.525 & 0.26 & 95 \\  
\ce{K3Cd5H12} & 13.468 & 0.00 & 81 \\  
\ce{K3Mg5H12} & 13.152 & 0.00 & 92 \\  
\ce{Li3Ca5H12} & 12.766 & 0.00 & 97 \\  
\ce{Li3Pd5H12} & 10.986 & 0.00 & 72 \\  
\ce{Li3Pt5H12} & 11.102 & 0.00 & 90 \\  
\ce{Li3Ta5H12} & 11.055 & 0.00 & 93 \\  
\ce{Li3Zn5H12} & 11.432 & 0.00 & 94 \\  
\ce{Dy3Li5H12} & 11.841 & 0.00 & 97 \\  
\ce{Er3Li5H12} & 11.734 & 0.00 & 100 \\  
\ce{Eu3Li5H12} & 12.370 & 0.00 & 82 \\  
\ce{Gd3Li5H12} & 11.970 & 0.00 & 92 \\  
\ce{Hf3Li5H12} & 11.085 & 0.00 & 93 \\  
\ce{Ho3Li5H12} & 11.786 & 0.00 & 98 \\  
\ce{Nd3Li5H12} & 12.276 & 0.00 & 92 \\  
\ce{Pr3Li5H12} & 12.355 & 0.00 & 84 \\  
\ce{Sc3Li5H12} & 11.218 & 0.00 & 93 \\  
\ce{Sm3Li5H12} & 12.113 & 0.00 & 93 \\  
\ce{Tb3Li5H12} & 11.900 & 0.00 & 95 \\  
\ce{Y3Li5H12} & 11.859 & 0.00 & 93 \\  
\ce{Zr3Li5H12} & 11.237 & 0.00 & 95 \\  
\ce{Lu3Rh5H12} & 11.393 & 0.23 & 94 \\  
\ce{Lu3Zr5H12} & 12.204 & 0.00 & 98 \\  
\ce{Mg3Pd5H12} & 11.305 & 0.00 & 91 \\  
\ce{Mg3Rh5H12} & 11.008 & 0.00 & 52 \\  
\ce{Na3Mg5H12} & 12.421 & 0.00 & 83 \\  
\ce{Na3Zn5H12} & 11.955 & 0.00 & 65 \\  
\ce{Eu3Na5H12} & 13.335 & 0.00 & 84 \\  
\ce{Rb3Cd5H12} & 13.772 & 0.00 & 94 \\  
\ce{Sc3Cu5H12} & 11.068 & 0.00 & 83 \\  
\ce{Sc3Pd5H12} & 11.356 & 0.00 & 74 \\  
\ce{Sc3Rh5H12} & 11.138 & 0.15 & 73 \\  
\ce{Sc3Ru5H12} & 11.010 & 0.00 & 71 \\  
\ce{Sc3Nb5H12} & 11.458 & 0.00 & 94 \\  
\ce{Sc3Ta5H12} & 11.380 & 0.00 & 83 \\  
\ce{Sc3Ti5H12} & 11.321 & 0.00 & 95 \\  
\ce{Ta3Mo5H12} & 11.027 & 0.10 & 98 \\  
\ce{Ta3Nb5H12} & 11.274 & 0.00 & 99 \\  
\ce{Nb3Ta5H12} & 11.256 & 0.00 & 80 \\  
\ce{Ti3Ta5H12} & 11.089 & 0.00 & 43 \\  
\ce{Tb3Rh5H12} & 11.597 & 0.23 & 99 \\  
\ce{Ti3Mo5H12} & 10.867 & 0.00 & 97 \\  
\ce{Ti3Nb5H12} & 11.148 & 0.00 & 59 \\  
\ce{Nb3Ti5H12} & 11.085 & 0.00 & 96 \\  
\ce{Tm3Rh5H12} & 11.453 & 0.24 & 95 \\  
\ce{Y3Pd5H12} & 11.824 & 0.00 & 99 \\  
\ce{Y3Rh5H12} & 11.581 & 0.27 & 93 \\  
\ce{Zr3Pd5H12} & 11.434 & 0.05 & 88 \\  
\ce{Zr3Nb5H12} & 11.543 & 0.00 & 63 \\  
\ce{Zr3Ta5H12} & 11.480 & 0.00 & 56 \\  
\ce{Zr3Ti5H12} & 11.378 & 0.00 & 63 \\  
\hline

%    \end{tabular}

\end{longtable}

\begin{figure*}[!h]
    \centering
    \begin{subfigure}[t]{0.9\columnwidth}
      \includegraphics[height=0.9\columnwidth]{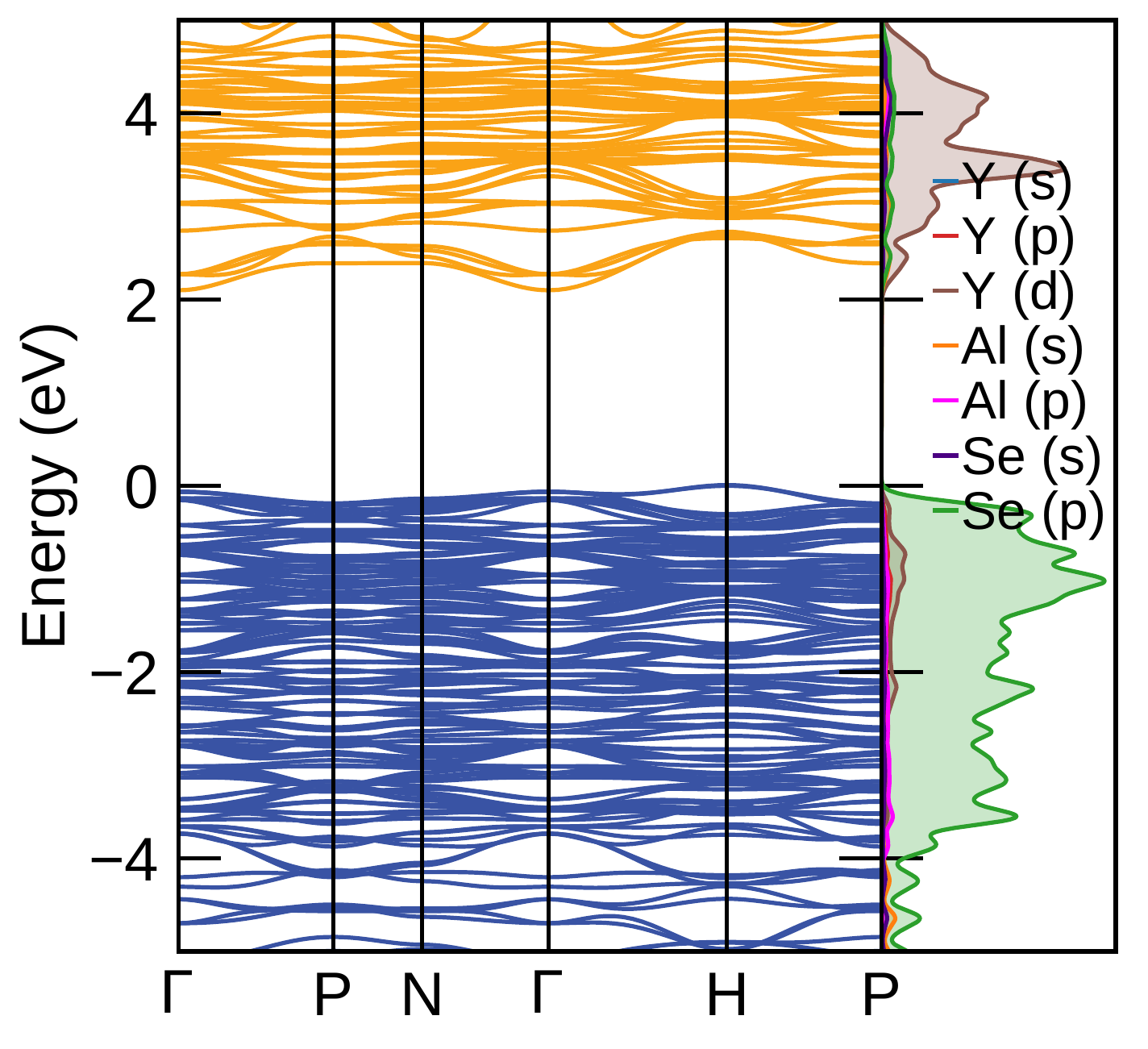} 
     \caption{\ce{Y3Al5Se12}}
     \end{subfigure}
    \begin{subfigure}[t]{0.9\columnwidth}
      \includegraphics[height=0.9\columnwidth]{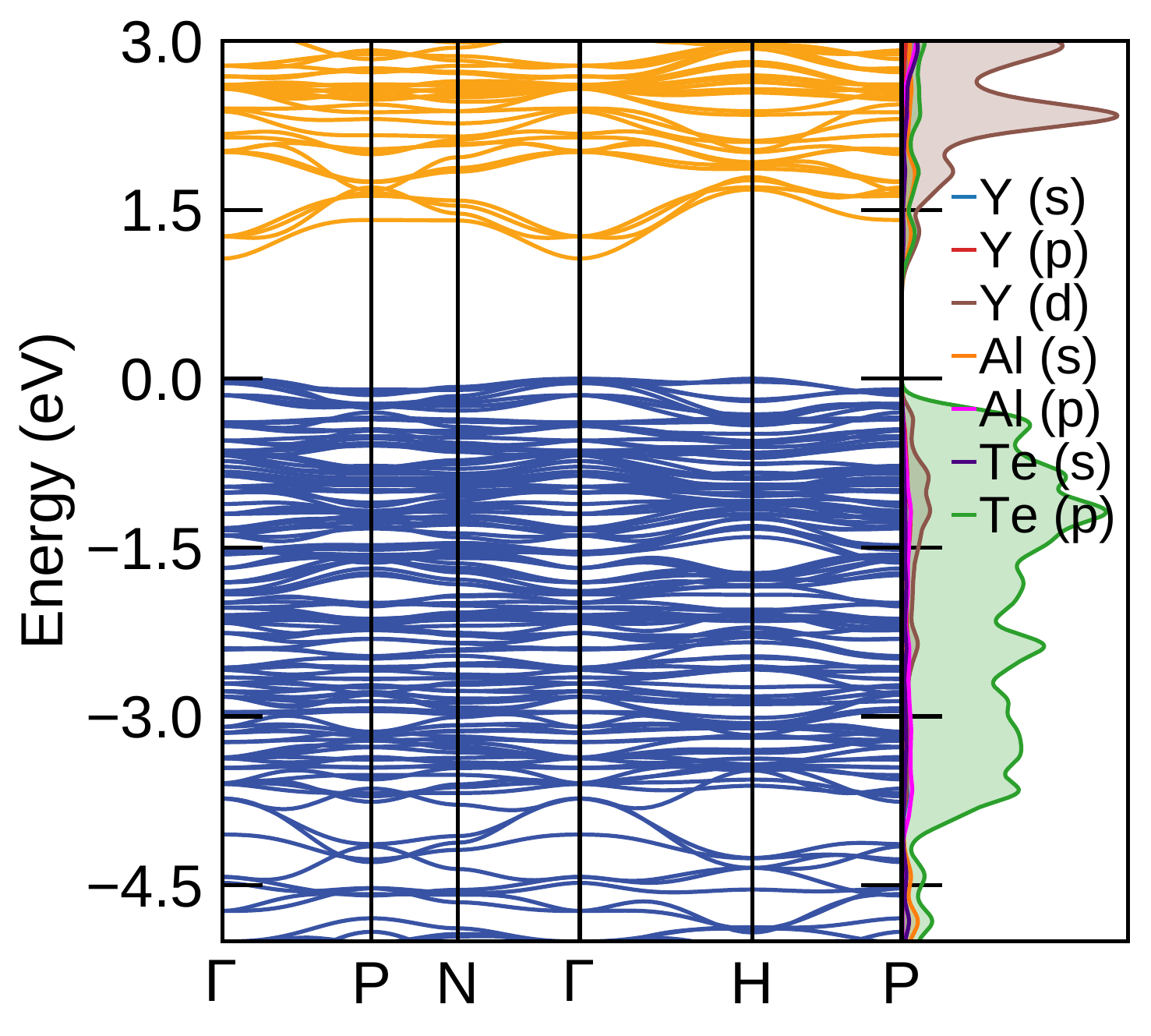}
     \caption{\ce{Y3Al5Te12}}
    \end{subfigure}
    \begin{subfigure}[t]{0.9\columnwidth}
      \includegraphics[height=0.9\columnwidth]{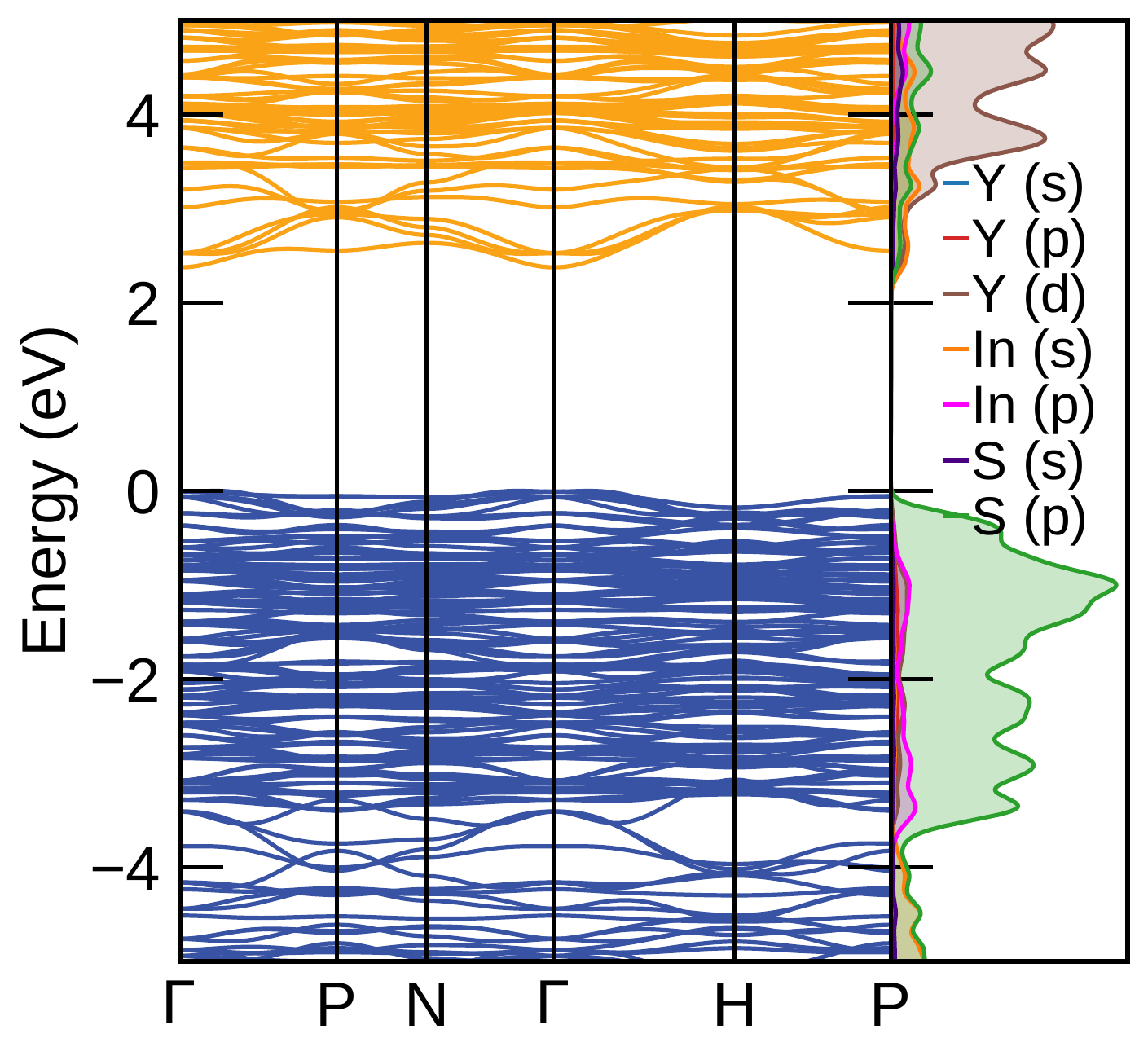} 
     \caption{\ce{Y3In5Se12}}
    \end{subfigure}
    \begin{subfigure}[t]{0.9\columnwidth}
      \includegraphics[height=0.9\columnwidth]{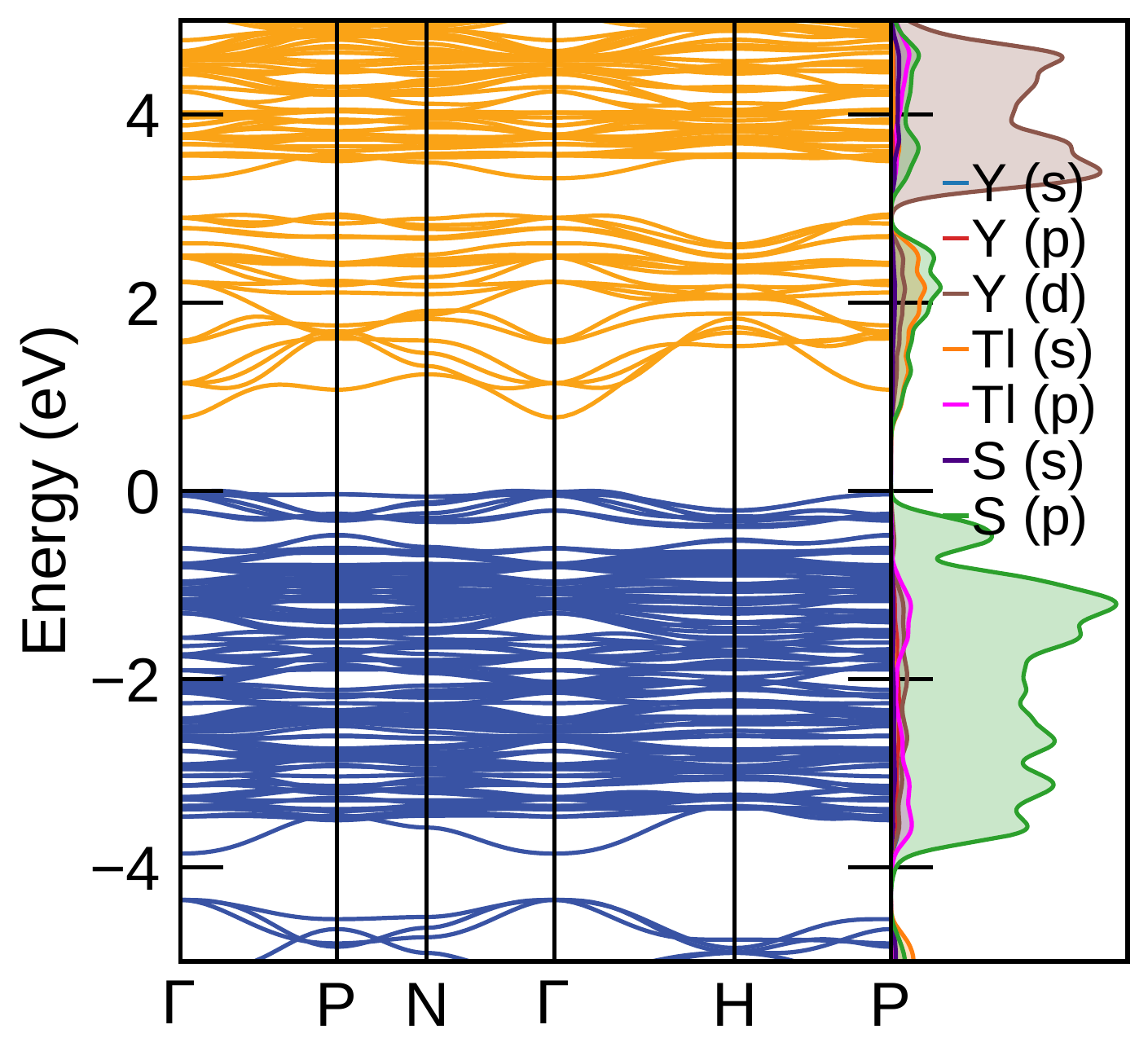}
    \caption{\ce{Y3Tl5Se12}}
    \end{subfigure}
    \begin{subfigure}[t]{0.9\columnwidth}
      \includegraphics[height=0.9\columnwidth]{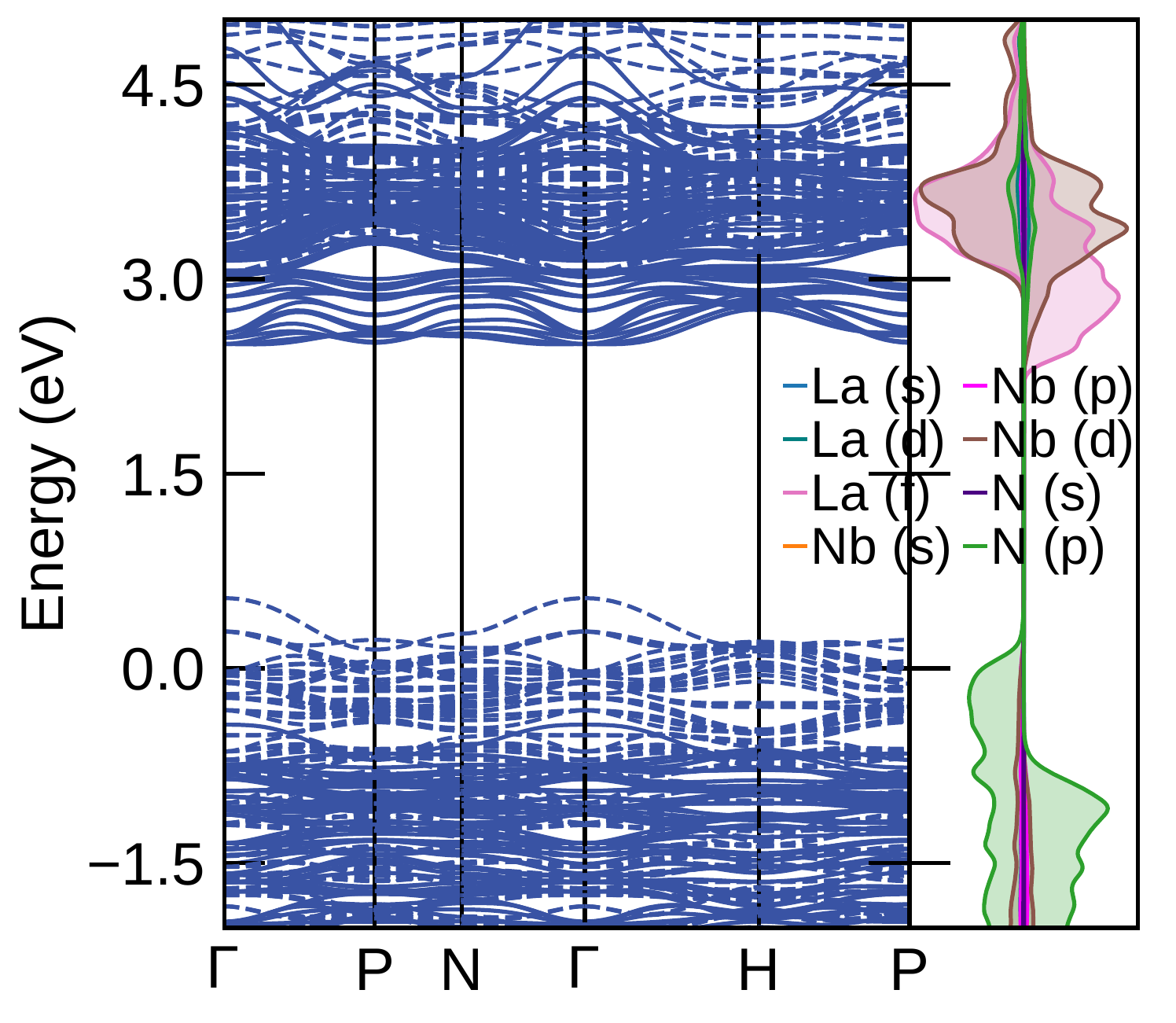} 
    \caption{\ce{La3Nb5N12}}
    \end{subfigure}
    \caption{The calculated mBJ electronic band structures of (a) \ce{Y3Al5Se12}, (b) \ce{Y3Al5Te12}, (c)\ce{Y3In5S12}, (d) \ce{Y3Tl5S12}, and (e)\ce{La3Nb5N12}. }
    \label{fig:sipp_bands}
\end{figure*}